\documentclass[12pt]{article} 

\usepackage{amsmath}
\usepackage{amssymb}
\usepackage[dvips]{graphicx}
\usepackage{natbib}
\bibpunct{(}{)}{;}{a}{}{,}

\title{Cavity opening by a giant planet in a
protoplanetary disc and effects on planetary migration.}

\author{\textsc{A. Crida}\footnote{correspondance should be sent to\,: Insitut f\"ur Astronomie und Astrophysik, Abt Computational Physics, Auf der Morgenstelle 10, D-72076 T\"ubingen, Germany ; e-mail\,: \texttt{crida@oca.eu}}, \textsc{A. Morbidelli}\\ \ \\
{O.C.A., B.P. 4229, 06304 Nice Cedex 4, France}}

\date{\small Accepted 2007 March 6. Received 2007 March 2\,; in original form 2007 January 24}

\begin{document}


\maketitle


\section*{Abstract}
We study the effect of a Jovian planet on the gas distribution of a
protoplanetary disc, using a new numerical scheme that allows us to
take into consideration the global evolution of the disc, down to an
arbitrarily small inner physical radius. We find that Jovian planets
do not open cavities in the inner part of the disc (i.e. interior to
their orbits) unless (a) the inner physical edge of the disc is close
to the planet's location or (b) the planet is much more massive than
the disc. In all other cases the planet simply opens a gap in the gas
density distribution, whose global profile is essentially unchanged
relative to the one that it would have if the planet were absent. We
recognize, though, that the dust distribution can be significantly
different from the gas distribution and that dust cavities might be
opened in some situations, even if the gas is still present in the
inner part of the disc.

Concerning the migration of the planet, we find that classical Type-II
migration (with speed proportional to the viscosity of the disc)
occurs only if the gap opened by the planet is deep and clean. If
there is still a significant amount of gas in the gap, the migration
of the planet is generally slower than the theoretical Type-II
migration rate. In some situations, migration can be stopped or even
reversed. We develop a simple model that reproduces satisfactorily the
migration rate observed in the simulations, for a wide range of disc
viscosities and planet masses and locations relative to the inner disc
edge. Our results are relevant for extra-solar planetary systems, as
they explain (a) why some hot Jupiters did not migrate all the
way down to their parent stars and (b) why the outermost of a pair of
resonant planets is typically the most massive one.

\paragraph{Key words\,:}
{\it (stars:)} planetary systems: formation --
{\it (stars:)} planetary systems: protoplanetary discs --
accretion, accretion discs --
Solar System: formation.

\vfill

\ 

\newpage

\section{Introduction}

The planet--disc interactions have been subject of an increased
interest since the discovery of the first exoplanet
\citep{MayorQueloz1995}. Indeed, the first extra solar planets
discovered were giant gaseous planets orbiting surprisingly close to
their parent star, so that they are called \emph{hot
Jupiters}. According to classical planetary formation models in
protoplanetary discs \citep{Pollack-etal-1996}, hot Jupiters could not
form where they currently orbit, because there was not enough solids
to build a massive core that would accrete a gaseous atmosphere (there
is much more solid material beyond the so-called `snow--line' where
water condensates into ice). In addition, gas heated by the star is
more difficult to capture for the solid core than cold gas. It has
been shown by \cite*{Bodenheimer-etal-2000} that in situ
formation of hot Jupiter is not impossible, but it is not the most
likely scenario. This suggests that the planets migrate in the disc,
via angular momentum exchanges with the gas.

In fact, the planet always exerts a positive torque on the part of the
disc outside its orbit, and a negative one on the part of the disc
inside its orbit \citep{GT79,LinPapaloizou1979,Ward1997}. Reciprocally
the disc exerts the opposite torques on the planet\,; if the disc
density profile is not perturbed by the planet, the sum of the torques
is not zero but results in the so-called \emph{differential Lindblad
Torque}, which is negative and responsible for the inward \emph{type~I
migration} \citep{Ward1997}. However, the torques exerted by the
planet on the disc tend to repel the gas away from the planetary
orbit\,; whether the internal stress in the disc is strong enough to
counterbalance this planetary torque and spread the gas into the void
regions determines whether the corotation region of the planet is
depleted or not, leading to the opening of a more or less clean
\emph{gap}. For more detail on the gap opening process, we refer to
\cite*{Crida-etal-2006} and references therein\,; the basic idea is
that the density profile evolves in response to the presence of the
planet, in order to achieve an equilibrium among three torques
exerted on a fluid element. These torques are due to the
gravity of the planet and to the viscosity and the pressure in the
gas\,; the two latter ones depend strongly on the gas density profile. The
stronger is the planetary torque (it is proportional to the planet
mass), the steeper has to be the density profile, and thus the deeper
is the gap.

If a clean gap is opened, the planet is locked into it because the
outer and the inner disc are repellent for the planet. Thus, the planet is
supposed to follow the viscous evolution of the disc \citep[accretion
on to the central star and viscous spreading, see][ hereafter
LP74]{LBP74}\,; this is the definition of \emph{type~II migration}.
However, \citet{Quillen-etal-2004} noticed that if the planet is more
massive than the inner disc, the outer disc is not able to push the
planet inward at the speed of the natural gaseous accretion on to the
star. Indeed, in that case the planet has more angular momentum to
lose than the unperturbed disc would have, so that it migrates more
slowly. Meanwhile, nothing prevents the inner disc from accreting on to
the star. Even more, \citet{Varniere-etal-2006} suggested that, as the
planet exerts a negative torque on the inner disc, it accelerates the
accretion of the latter on to the primary. Consequently, the inner disc
accretes on to the central star faster than it would naturally. This
makes the inner disc disappear before the planet migrates inward. The
result is the opening of a \emph{cavity}\,: a great depletion of the
disc from the star to the planet's orbit.

Once a cavity is opened, the outer disc still tends to accrete on to
the central star and to push the planet inward, while the planet can
not take angular momentum from the inner disc anymore. A big mass of
the planet with respect to the disc may slow down this process, but
the final result should ineluctably be the inward type~II migration of
the planet toward the star. Reality, however, is not so simple. In
fact,  in this work we will show that, under
some conditions on the gas disc viscosity, the orbit of the planet is
not `gas-proof' and viscous accretion can happen without forcing the
migration of the planet. 

The cavity opening process and its feedback on planet migration are
not easy to study and quantify with the help of numerical
simulations. Indeed, the computation of the inner disc evolution is
prohibitive with classical hydro-codes. Thus,
\citet{Quillen-etal-2004} only made the simulation of a planet
orbiting inside a pre-existing cavity --\,showing that it did not
migrate inward very fast\,-- and \citet{Varniere-etal-2006} performed
only a few simulations showing the opening of a cavity by a
Jupiter-mass planet in a low density disc. The cavity opening appeared
to be more rapid than the viscous time-scale, confirming the idea that
the planet `helps' the accretion of the inner disc on to the
star. However, \citet{Varniere-etal-2006} did not provide an
exploration of the parameter space, and could not study the planetary
migration because the planet was about 100 times more massive than the
disc. With the help of a new numerical scheme that simulates
efficiently and realistically the inner disc evolution
\cite*{Crida-etal-2007}, we can perform such a study for the first
time. This is the main scope of this paper. This analysis is
particularly interesting because observations suggest that some
protoplanetary discs most likely host cavities. Understanding what
kind of cavity a planet can open in terms of size, shape and depth,
in which conditions and with which life-time, may provide some keys to
interpret these observations. We also stress that the ALMA
interferometer will enable us to make precise images of the inner part
of protoplanetary discs. In this frame, it is relevant to study the
features resulting from planet--disc interactions that might be
observed.

In Section~\ref{sec:observations}, we will shortly review the
observations of protoplanetary discs that seem to host cavities. After
a brief description of the algorithm used for our numerical
simulations (Section~\ref{sec:numerics}), we present the results in
Section~\ref{sec:results} in terms of cavity opening and
migration. These numerical experiments are interpreted in
Sections~\ref{sec:interpretation_dp} and~\ref{sec:interpretation_migr}
with a semi-analytical model. We discuss in particular the
survivability of a cavity opened by a planet with regard to type~II
migration, and we find that the corotation torque can prevent this
inward migration. In Section~\ref{sec:dust} we discuss the difference
between gas cavities and dust cavities. Finally, we conclude and
discuss the implications of this work in Section~\ref{sec:conclu}.

\section{Evidences for cavities in protoplanetary discs.}

\label{sec:observations}

Discs of gas and dust around young stars are heated by the star and by
internal viscous dissipation. Consequently, the dust becomes an
infra-red source. In the spectral energy distribution (SED) of the
star, an infra-red excess reveals the presence of such a
disc. The near-infra-red (NIR) emission corresponds to the hottest part of
the disc, which is the closest to the central star, while the
mid-infra-red emission is due to colder dust, corresponding to the
outer part of the disc. Consequently, a lack of infra-red excess in the NIR
part of the spectrum can be interpreted as the presence of an
inner cavity in the disc \citep{Beckwith1999}. Such a cavity
discovered by SED analysis is sometimes called a \emph{spectral hole}.

One of the most clear examples of spectral hole is given in fig.~2 in
\citet{Forrest-etal-2004} for the T-Tauri star CoKu Tau/4 observed
with the Spitzer telescope. The size of the spectral hole is such that
there seems to be no dust at temperature larger than 123K. The authors
convincingly deduce from this SED that dust grains in the disc of CoKu
Tau/4 are excluded from the innermost 10 au\,; this makes a remarkable
dust cavity. The question whether this cavity could be opened and
maintained by a planet has been addressed in
\citet{Quillen-etal-2004}. They suppose that a planet opened a gap in
the disc at 10 au, and that the inner disc subsequently accreted on to
the star. Then the age of the system has to be bigger than the viscous
time-scale of accretion on to the star of the inner disc. This gives a
lower limit for the viscosity. Given the viscosity, a minimum planet
mass is required to open a gap. For the planet not to migrate
immediately after the depletion of the inner disc, the planet has to
be more massive than the outer disc, so that the angular momentum
taken by the latter from the planet is a negligible fraction of the
angular momentum of the planet. From these considerations,
\citeauthor{Quillen-etal-2004} performed a simulation with a planet of
mass $3\times 10^{-4}M_*$ on a circular orbit at 10 au in a disc with
Reynolds number $10^5$ and density $\Sigma(r)=
10^{-6}(1.1r_p/r)\,M_*{\rm au}^{-2}$ for $r>1.1r_p$ and 100 times
smaller for smaller radii. The planet efficiently maintained the
cavity for the duration of the simulation (100 orbits). However, they
did not make a simulation of the process of cavity opening.

The T-Tauri stars TW Hya, DM Tau and GM Aur also present a spectral
hole. To interpret their SEDs, \citet{Calvet-etal-2002,
Calvet-etal-2005} used the following model. They divided the disc into
3 components\,: an optically thick outer disc, its inner edge
(represented as a wall directly exposed to stellar radiations), and an
optically thin inner region. Adjusting the free parameters (the wall
radius and height, the outer disc scale height, its mass and
viscosity) they obtained a satisfactory fit of the SED of these three
objects. They concluded that TW Hya and DM Tau present a 3--4 au wide
cavity, while GM Aur should have the truncation radius of the outer
disc located at 24 au from the star. Previous work on GM Aur
\citep{Bergin-etal-2004} based on different data, concluded that the
cavity is 6-au wide. Such a cavity could be maintained by a 1.7
Jupiter mass planet on a fixed circular orbit at 2.5 au
\citep{Rice-etal-2003}.

\cite*{Dullemond-etal-2001} proposed a different model for
the SED of Herbig Ae stars that does not involve the presence of a
planet.  In their model, the inner part of
the disc is removed, so that the inner rim is directly exposed to
stellar radiations. It is thus heated a lot, and gets puffed up. This
has two consequences\,: first, this rim is a strong IR source, leading
to a peak at 2--3 $\mu m$ in the SED\,; second, the disc behind the
puffed up inner rim is in the shadow. Thus, this region remains cold
and does nearly not radiate in IR. If the disc is flared, however, the
outermost part of the disc can be lightened by the star, and then
becomes an IR source. In that case, the SED presents a gap between the
peak at 2--3 $\mu m$ due to the hot rim and the emission of the outer
disc at larger wavelength, because it is colder. This is in good
agreement with the SED of the Herbig Ae star AB Aur, with a rim of height
0.1 au at 0.52 au (which corresponds to the dust sublimation radius).

More recently, interferometric observations at Plateau de Bure
Interferometer clearly showed a remarkable dust depletion in the inner
50 au of the disc of LkCa15 \citep{Pietu-etal-2006}. The
possibility that a massive planet or a brown dwarf 
is responsible for this is work in
progress and will be addressed in a forthcoming article.

Thus, cavities are relatively common features in protoplanetary
discs. It is not easy to constrain their characteristics, as sometimes
different dust distributions can give indiscernible SEDs, but the
evidence for cavities in protoplanetary discs is quite
strong.

\section{Numerical simulations setup.}

\label{sec:numerics}

Classical hydro-codes for the simulation of planet--disc interactions
represent the disc with a 2D polar grid extending between an inner and
an outer radius ($R_{\rm in}^{\rm 2D}$ and $R_{\rm out}^{\rm
2D}$). The grid covers the planetary region and, provided that the
resolution is good enough, satisfactorily describes the local
planet--disc interactions. The grid cannot reasonably cover the full
disc, for the following reasons. First, a prohibitive number of cells
would be required to extend the grid from the planet region (a few
astronomical units from the star) to the physical outer edge of the
disc (often located at several hundreds of au from the star). Second,
and more important, the simulation of the inner part of the disc
requires a shortage of the time-step and an increase of CPU time
roughly proportional to $\Omega_{\rm max}^{-1} \propto {R_{\rm
in}^{\rm 2D}}^{\,-3/2}$ in most codes, where $\Omega$ is the angular
velocity. Because the grid covers only a portion of the disc, the
global evolution of the disc cannot be realistically described. Yet,
it is the global evolution of the disc that governs type~II
migration. Thus, the accuracy of the type~II migration observed in
simulations can be questioned. Another point related to the global
evolution of the disc is the evolution of its innermost part and its
accretion on to the star, that is the opening of a
cavity. Consequently, it seems that classical hydro-codes are not
adapted for the study of these two problems on which we wish to
focus. We developed an improvement to the classical algorithm, that we
briefly present below.

\subsection{Code description}

The code used for this work is based on the hydro-code FARGO
\citep{FARGO,FARGO2}. To describe correctly the evolution of the disc
outside the 2D grid, we add a 1D grid extended all over the physical
disc, from its inner edge (the radius of the central star or the
X-wind truncation radius) to its outer edge. This non-azimuthally
resolved grid is coupled to the 2D grid at the boundaries of the
latter in a conservative way. Thanks to this coupling, the disc
perturbations due to the planet --\,computed locally in the 2D grid in
FARGO\,-- influence the global evolution described in the 1D grid\,;
this evolution in turn affects the computation in the 2D grid as it
provides realistic, time-evolving boundary conditions for the
latter. This numerical scheme, described in \citet{Crida-etal-2007},
enables us to simulate the type~II migration with an excellent
reliability as well as the accretion of the inner disc on to the star,
for a negligible additional computing cost with respect to a classical
2D hydro-code.

\subsection{Units and parameters}

\label{sec:units}

The central star mass is set equal to $M_\odot$, which is our mass
unit. As length unit, we adopt the astronomical unit (au), and we set
for simplicity the gravitational constant $G=1$ (so that one
year lasts $2\pi$ time units).

The initial density profile corresponds to a disc that evolved for
some time under the effect of its own viscosity and was provided by
\citet{GuillotHueso2006}. It can be approximated by\,:
\begin{equation}
\Sigma(r)\approx 10^{-5}\,\exp(-r^2/1320).
\label{eq:sigma_init}
\end{equation}
This initial density corresponds to the Minimal Mass Solar Nebula
\citep{Hayashi1981} at the location of Jupiter (5.2 au). The
aspect ratio $H/r$ is constant in space and time\,; its default value is
$0.05$. These parameters (density and aspect ratio) will be changed
in some test runs to study their influence on the results. The sound
speed is $c_s=H\Omega$, where $H$ is the height of the disc. The
equation of state is isothermal (the pressure is $P=\rho c_s^2$, where
$\rho$ is the volume density, so that it becomes $P=\Sigma c_s^2$ in
our 2D formalism, after integration on the disc's scale height).

If not specified otherwise, in most of the runs the 2D grid extends
radially from $R_{\rm in}^{\rm 2D}=1.75$ to $R_{\rm out}^{\rm 2D}=15.$
au and is divided in $N_r=165$ elementary rings, themselves divided
into $N_s=320$ sectors. The 1D grid extends radially from $0.58333$ to
$100$ au, over $N_r^{\rm 1D}=1193$ elementary rings.

\section{Results}

\label{sec:results}

We computed a few simulations with a Jupiter mass planet
($q=M_p/M_*=10^{-3}$) initially placed on a circular orbit at $r_p=5$
au in a disc with different viscosities. The kinematic viscosity of
the gas, $\nu$, is constant in space and time. Its values in the
simulations are such that the Reynolds number at the initial location
of the planet ($\mathcal{R}={r_p}^2\Omega_p/\nu$) goes from $10^{3.8}$
to $10^{6}$.

\subsection{Density profile and cavity opening}

\label{subsec:results_dpco}
\begin{figure}
\includegraphics[width=0.7\columnwidth, angle=270]{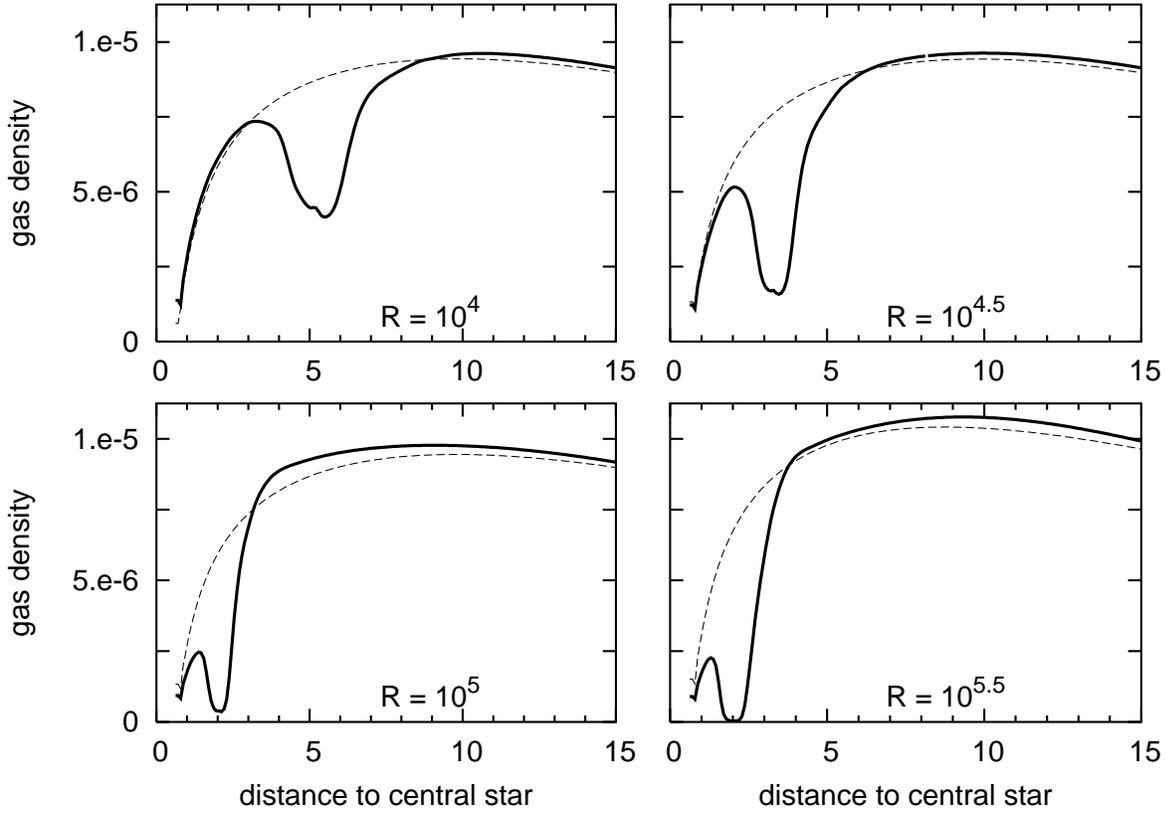}
\caption{Bold lines\,: Surface density profile at $t_\nu/2$, for
different Reynolds number, labelled on the corresponding panel.  For
$\mathcal{R}=10^{5.5}$, the time is $t_\nu/4$.
Thin dashed lines\,: surface density profile at the same times for a
similar disc with no planet. The units of mass and length are
those specified in Section~\ref{sec:units}\,: solar mass and astronomical
unit.}
\label{fig:profils}
\end{figure}

Fig.~\ref{fig:profils} shows with bold lines the density profiles for
four simulations after half a viscous time ($t_\nu=\nu\,r_p^{\,-2}$), or
only a quarter of viscous time in the least viscous case. The depth of
the gap opened by the planet strongly depends on the Reynolds
number. This is not surprising, as it is well-known that viscosity
plays against gap opening \citep[see for instance ][ and references
therein]{Crida-etal-2006}. The gap is centred on the planetary orbit,
which is not any more at $5$ au because the planet has migrated.

Concerning the opening of a cavity, we see that the density in the
inner disc is significantly smaller than in the outer disc. In
particular the least viscous cases show an inner disc with very low
density, so that a sort of cavity is formed\,: in fact, the density is
about $5$ times smaller for $r<2.5$ than for $r>3$. The profile
clearly shows a wall at $r\approx 2.5$, beyond which the disc profile
is about flat while the density is negligible at the base of the wall.

The thin dotted lines show the density profile obtained in similar
simulations, at the same time, in the same conditions, but without the
planet. The density also decreases in the innermost region of the
disc. So, it seems that the observed depletion of the inner disc is
not fully caused by the planet, as will be discussed in
Section~\ref{sec:interpretation_dp}.

\subsection{Migration rate}

\begin{figure}
\includegraphics[width=0.7\columnwidth, angle=270]{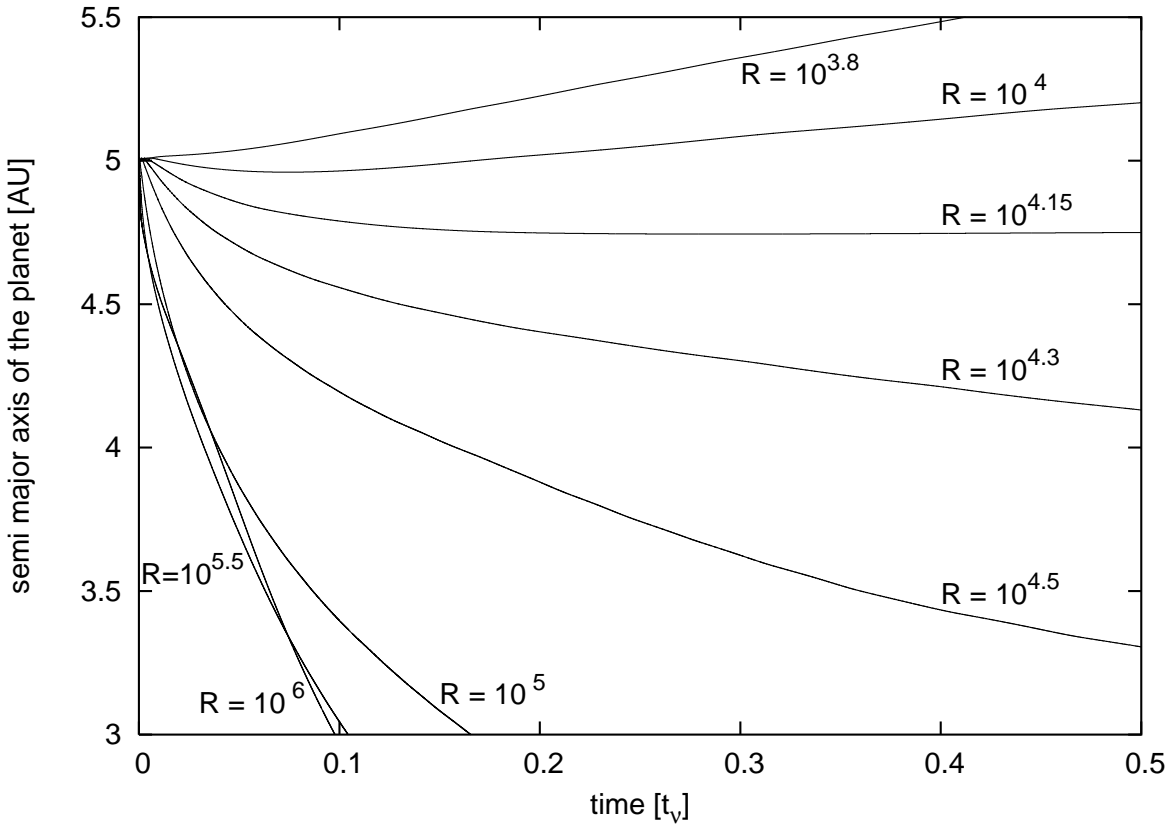}
\caption{Migration of a Jupiter mass planet for different Reynolds
numbers of the disc.}
\label{fig:Migr_tvisq}
\end{figure}

Fig.~\ref{fig:Migr_tvisq} shows the evolution of the semi major axis
of the planet with time, in units of the viscous time $t_\nu$. In
proper type~II migration, the migration rate is proportional to the
viscosity and thus, in this time unit, it should be independent of
$\mathcal{R}$. This is indeed the case for $\mathcal{R} \geqslant
10^{5}$\,: the three curves are almost linear and overlap, at least in
the first part of the evolution. However, for higher viscosities, a
very different behaviour is observed. When $\mathcal{R}$ decreases
below $10^{5}$, the migration rate becomes slower than expected. As
viscosity increases, the planets migrates inward more and more slowly
with respect to the viscous time, and the migration is even stopped
for $\mathcal{R}=10^{4.15}$. We will refer to this case hereafter as
the {\it stationary case}. For larger viscosities, the migration is
reversed and the planet moves outward.

\begin{figure}
\includegraphics[width=0.7\columnwidth, angle=270]{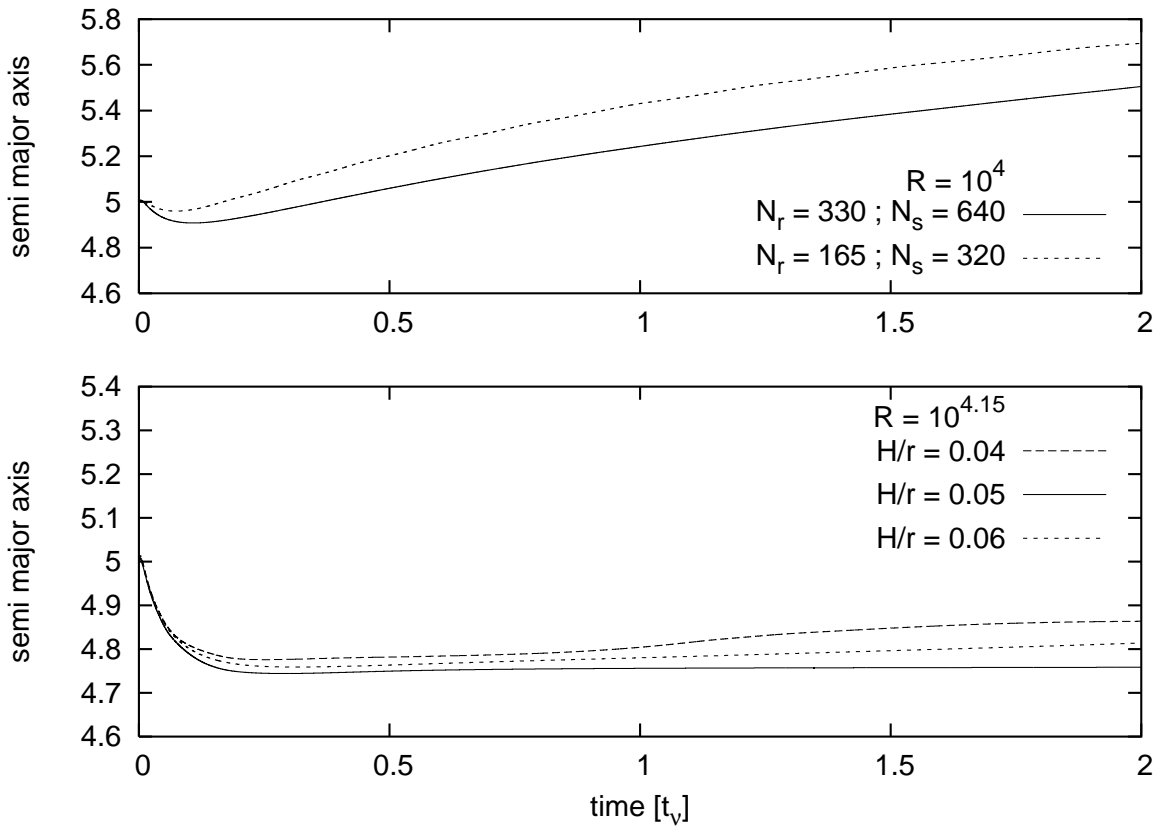} 
\caption{Top panel\,: Outward migration of a Jupiter mass planet for
$\mathcal{R}=10^4$ and two different resolutions. Bottom panel\,:
Quasi stationary evolution of a Jupiter mass planet for
$\mathcal{R}=10^{4.15}$ and 3 different aspect ratios.}
\label{fig:migr-params}
\end{figure}

This result is particularly surprising and new.  This outward
migration is not an effect of the resolution of the grid\,; we have
recomputed the simulation with $\mathcal{R}=10^4$, using $N_r=330$,
$N_s=640$\,: the planet also migrates outward, at about the same speed
(see top panel in Fig.~\ref{fig:migr-params}). With this high
resolution, the size of a cell around the planet is about a tenth of
the Hill Radius in radius and a seventh of the Hill radius in azimuth,
so that the Hill sphere of the planet is covered by about 220
cells. This is largely sufficient; in particular the corotation zone
of the planet, which plays a crucial role as will be shown further, is
correctly simulated. We also investigated the effect of the aspect
ratio on the migration rate, from the stationary case. It appears that
this parameter only plays a marginal role (see bottom panel in
Fig.~\ref{fig:migr-params}).

\begin{figure}
\centering
\includegraphics[width=0.6\columnwidth, angle=270]{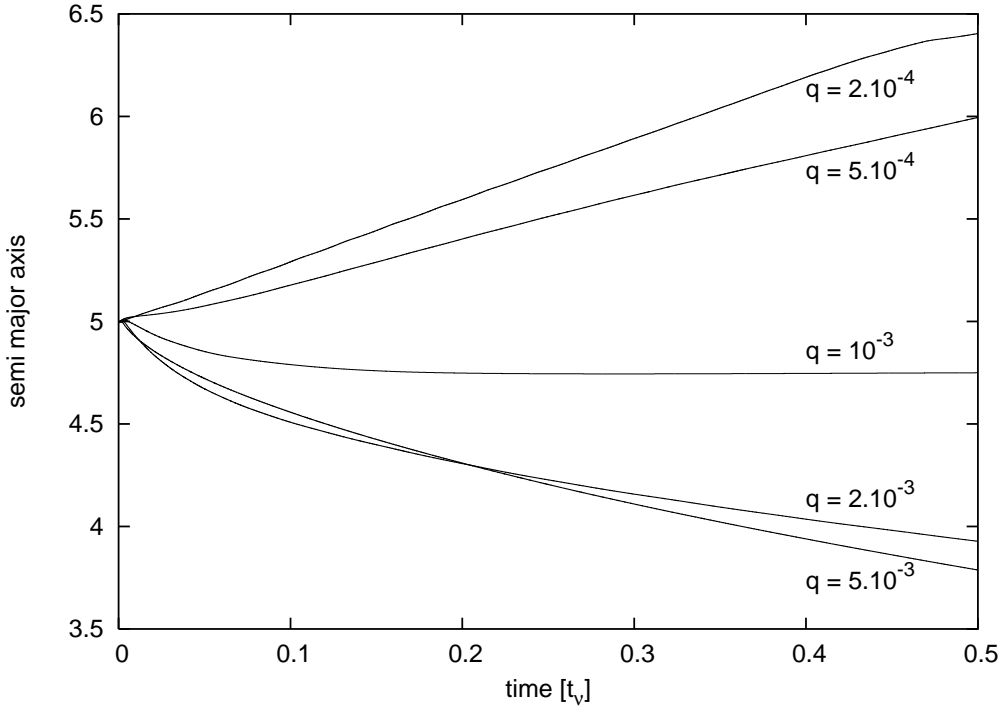}
\caption{Migration of giant planets of different masses in a disc with
$\mathcal{R}=10^{4.15}$ and $H/r=0.05$.}
\label{fig:migr-mass}
\end{figure}

One may wonder if the stationary case observed for
$\mathcal{R}=10^{4.15}$ for a Jupiter mass planet is a feature valid
for any planet mass. Fig.~\ref{fig:migr-mass} shows that it is not the
case. The more massive is the planet, the faster it migrates inward,
approaching the classical type~II regime. For planets lighter than
Jupiter, the migration is directed outward.  However, we will show in
Section~\ref{sec:interpretation_migr} that if the viscosity is
independent of radius, there is a stationary orbit further from the
star which these low mass planets tend to, asymptotically.

\section{Modelling the density profile}

\label{sec:interpretation_dp}

In this Section we build a model to understand the shaping of the
disc density profiles observed in Fig.~\ref{fig:profils}. 
Our approach is done in three steps. First, we review the global
viscous evolution of a disc that is not perturbed by any planet; then, 
we discuss how a gap is opened by a massive planet  
and finally we analyse in which
situations the resulting density profile may show a significant depletion
(a `cavity') in the full inner part of the disc, up to the planetary orbit.  

\subsection{Global disc's viscous evolution}

The natural viscous evolution of a gaseous disc is described in
LP74. Using their partial differential equation~(14), one can compute, with
a step by step integration, the evolution of any initial profile. If
the initial profile has a Gaussian shape
($\Sigma(r)=\Sigma_0\exp(-ar^2)$, where $a$ is an arbitrary constant),
there is an explicit solution, given by equations~(18) and (18') of LP74.
The initial shape is conserved, the surface density being\,:
$$\Sigma(r) = \Sigma_0 T^{-5/4}\exp\left(-\frac{ar^2}{T}\right)\ ,$$
where $T=12a\nu t +1$. 

\begin{figure}
\centering
\includegraphics[width=0.6\columnwidth, angle=270]{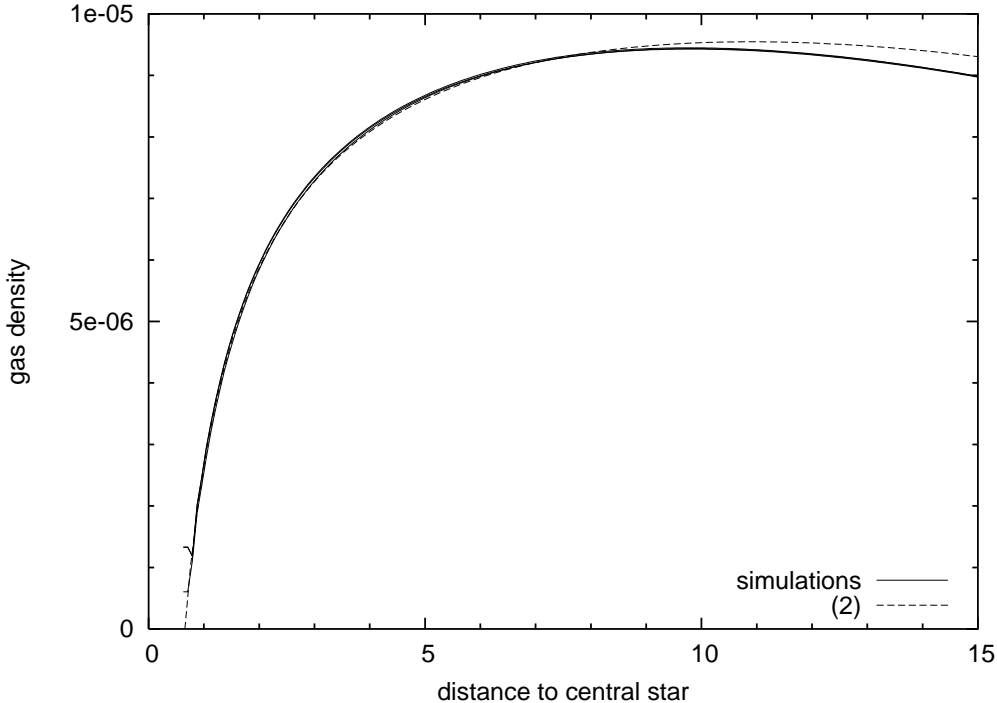}
\caption{Surface density profile of an unperturbed disc at
$t=t_\nu/2$. Plain lines\,: profiles from simulations with
$\mathcal{R}=10^{\,4}$, $10^{4.5}$, and $10^{\,5}$. Dashed line\,:
profile from equation~\eqref{eq:LBP}.}
\label{fig:LBP}
\end{figure}

This solution is valid only for a disc extending between $0$ and
infinity in radius. However, the inner radius of the disc, $R_{\rm
inf}$, is never $0$\,; it is at least the radius of the central star,
most likely the corotation radius at a few tenth of au where the lines
of the magnetic field reconnect (the so-called X-point;
\citet{Shu-et-al-1977}). It could even be bigger, in the case of Jet
Emitting Disc (JED)\,: indeed, at the base of the jet, which may be
several-au-large (\citet{Bacciotti-etal-2003};
\citet{Coffey-etal-2004}), accretion is dominated by the torque
exerted by the jet and is much larger than the standard accretion\,;
so, the outer radius of the jet could be considered as the inner open
boundary of the standard disc. The explicit solution of the LP74
equations with a finite inner edge of the disc comes from
equation~(25) of LP74. With our units ($G=M_*=1$), it gives\,:
\begin{equation}
\Sigma_{\rm LP74}(r,t)=\Sigma_0T^{-5/4}\left(\frac{\sqrt{r}-\sqrt{R_{\rm inf}}}{\sqrt{r}}\right)\exp\left(-\frac{ar^2}{T}\right)\ .
\label{eq:LBP}
\end{equation}
Thus the density starts from $0$ at $r=R_{\rm inf}$ and grows with $r$
until $r\gg R_{\rm inf}$, where a classical Gaussian shape is
reached. This kind of profile is illustrated by the thin dashed lines
in Fig.~\ref{fig:profils}. Fig.~\ref{fig:LBP} shows the agreement
between the disc profiles obtained for various viscosities in
numerical simulations and the theoretical profile given by
equation~\eqref{eq:LBP} with $\Sigma_0=1.224\times 10^{-5}$ and
$a=1/1320$ at $t_\nu/2$, (\textit{id est} $T=1.114$). All profiles
overlap with each other. This shows that, once the time is
renormalised relative to the viscous time, the evolution of the disc
density distribution is independent of the Reynolds number. Also, it
shows the excellent agreement between equation~\eqref{eq:LBP} and the
numerical solution achieved with our simulation scheme.  The little
discrepancy visible in Fig.~\ref{fig:LBP} comes from the fact that the
original profile in the simulations is not given by
equation~\eqref{eq:LBP} at $T=1$ but by
equation~\eqref{eq:sigma_init}.

\subsection{Gap opening in an evolving disc}

\label{subsec:Gap-opening}

Equation~(14) of \citet{Crida-etal-2006} provides a way of
computing semi-analytically the density profile of a gap for any disc
and planet parameters. This equation comes from the equilibrium
between the torques due to the gravity of the planet, the viscosity
and the pressure in the disc. It applies outside the corotation zone
of the planet, which approximately extends on each side of the orbit
over a width $x_s^+\approx x_s^-\approx 2R_H$, where $R_H$ is the Hill
radius of the planet.  the equation reads\,:
\begin{equation}
\frac{R_H}{\Sigma}\frac{{\rm d}\Sigma}{{\rm d}r} =
\cfrac{0.35q^2\mathcal{R}{r_p}^3r\Delta^{-4}{\rm sgn}(\Delta)-\cfrac{3}{4}\cfrac{\Omega}{\Omega_p}}{\left(\cfrac{H}{r}\right)^{\!2}\cfrac{r}{r_p}\mathcal{R}a''+\cfrac{3}{2}\cfrac{r}{R_H}\cfrac{\Omega}{\Omega_p}}\ ,
\label{eq:Crida}
\end{equation}
where $a''=\frac18\left|\frac{\Delta}{R_H}\right|^{-1.2} +
200\left|\frac{\Delta}{R_H}\right|^{-10}$ and $\Delta = r-r_p$.

Notice that a boundary condition is needed to solve this differential
equation with a step by step procedure. The solution of this equation
corresponds to the gap profile $\Sigma(r)$ in an ideal disc in steady
state, under the influence of a non-migrating planet. Without the
presence of the planet, the unperturbed profile of such a disc would
be proportional to $r^{-1/2}$\ because $\nu$ is assumed to be constant
over $r$. Thus, to compute the profile in the inner part of the disc,
we start from $\Sigma=\Sigma_0/\sqrt{R_{\rm inf}}$ at $r=R_{\rm inf}$
and we compute the density step by step until $r^{-}=r_p-x_s^-$. For
the outer disc, we start with $\Sigma=\Sigma_0/\sqrt{R_{\rm sup}}$ at
$r=R_{\rm sup}$ and follow equation~\eqref{eq:Crida} down to
$r^{+}=r_p+x_s^+$. Unfortunately, $\Sigma(r^-)/\sqrt{r^-}$ is not
necessarily equal to $\Sigma(r^+)/\sqrt{r^+}$. Thus, we choose the
lowest of these two values, say the one at $r^-$, and adjust the outer
edge of the gap at $r^++\delta$, with $\delta$ such that
$\Sigma(r^++\delta)/\sqrt{r^++\delta}=\Sigma(r^-)/\sqrt{r^-}$.  As
shown in \citet{Crida-etal-2006}, this procedure gives a satisfactory
approximation of the gap depth.

The use of equation~\eqref{eq:Crida} as explained above provides the
gap profile in a disc whose unperturbed profile is $\Sigma(r)=\Sigma_0
r^{-1/2}$. Thus, multiplying this gap profile by $\sqrt{r}/\Sigma_0$
gives the gap profile in terms of fraction of the unperturbed
profile. Let us denote $\sigma(r)$ this \emph{fractional profile}.  We
think that it is reasonable to assume that the fractional profile is
independent of the unperturbed profile. This is supported by
Fig.~\ref{fig:profils}, which shows that the planet opens a gap in the
natural profile of the disc, without essentially changing
it. Consequently, as a simple model of the gap profile in an evolving
disc, we suggest that the unperturbed profile given by
equation~\eqref{eq:LBP} can be multiplied by the fractional profile,
obtaining\,:
\begin{equation}
\Sigma(r)=\sigma(r)\Sigma_{\rm LP74}(r,t)\ .
\label{eq:gap-model}
\end{equation}

\begin{figure}
\centering
\includegraphics[width=0.6\columnwidth, angle=270]{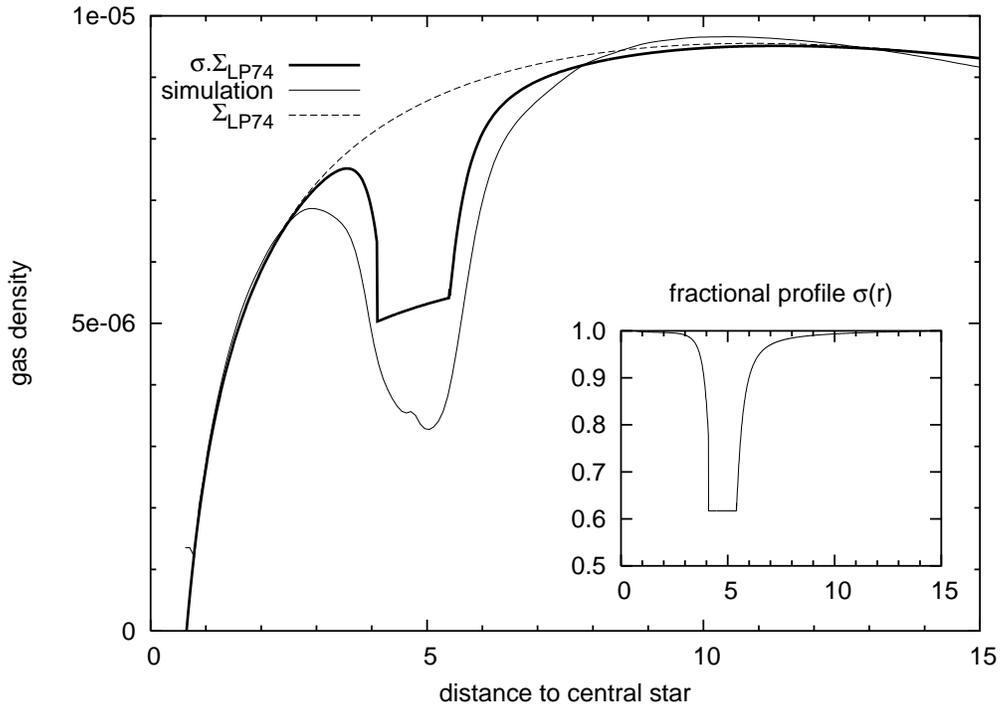}
\caption{Gap profile in an evolving disc with
$\mathcal{R}=10^{4.15}$\,: the thin line results from numerical
simulation while the bold line comes from our simple model. Dashed
line\,: the unperturbed profile $\Sigma_{\rm LP74}(r,t_\nu/2)$ from
equation~\eqref{eq:LBP}. Secondary panel\,: `fraction profile'
$\sigma(r)$ of the gap deduced from equation~\eqref{eq:Crida}.}
\label{fig:model}
\end{figure}

Fig.~\ref{fig:model} displays this model in the
$\mathcal{R}=10^{4.15}$ case. The secondary panel shows the fraction
profile $\sigma(r)$. In the main plot, the dashed line shows the
unperturbed profile $\Sigma_{\rm LP74}(r,t)$ at $t=t_\nu/2$.  The gap
profiles from the numerical simulation at the same time (thin line)
and from equation~\eqref{eq:gap-model} (bold line) are compared. The
agreement is quite satisfactory, in particular if one keeps in mind
that we only have a semi-analytic estimate of the gap profile. Indeed,
as discussed in \citet{Crida-etal-2006}, equation~\eqref{eq:Crida}
provides gap profiles that are a bit too narrow and shallow for
$\mathcal{R}\leqslant 10^5$ (while it provides slightly too wide and
deep gaps for $\mathcal{R}>10^{5.5}$). This is observed here as well.
However, the gap edges in the model correspond quite well to the ones
from the simulation in terms of position and slope. In addition, the
profiles in the inner and outer disc nearly match those observed in
the simulation.

We present here only this case, corresponding to a non-migrating planet.  In
the cases of a migrating planet, the situation is similar. 
This shows that this simple idea of multiplying the unperturbed profile
by a schematic gap profile is valid in first approximation. As a matter of
fact, a slight depletion of the inner disc with respect to the unperturbed
profile is visible for cases with inward migrating planets. We will explain
this in Section~\ref{sec:depletion}, after having developed a
model that reproduces the migration rates.

\subsection{Cavity opening}

\label{subsec:Cavity-opening}

As we have seen above the disc density profile can be described
effectively as the product of the 
fractional gap profile with the profile described
by equation~\eqref{eq:LBP}. This equation 
gives profiles which are proportional to $\left(1-(r/R_{\rm
inf})^{-1/2}\right)\exp(-ar^2/T)$. The first term is about $1$ for
$r>30 R_{\rm inf}$, so that the profile at large radius is about
Gaussian. However, for $r/R_{\rm inf}<5$, the first term shapes the
profile growing from 0 at $r=R_{\rm inf}$, with a slope proportional
to $(r/R_{\rm inf})^{-3/2}$. Thus, it seems that $R_{\rm inf}$ is a key
parameter for the profile of the innermost part of an evolving disc
\citep[see also ][]{LubowDAngelo2006}.

\begin{figure}
\includegraphics[width=0.7\columnwidth, angle=270]{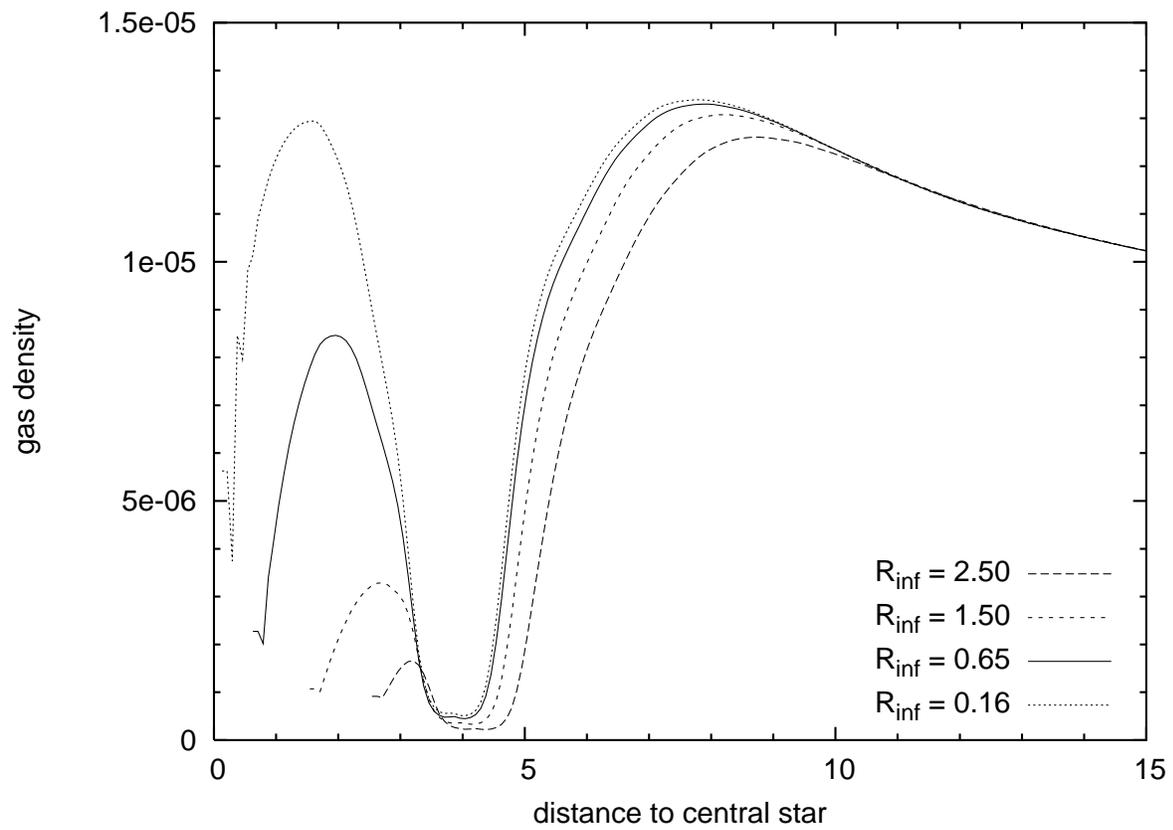}
\caption{Density profiles obtained in numerical simulations 
at $t=t_\nu/10$ for discs with $\mathcal{R}=10^5$ and 
different values of $R_{\rm inf}$.}
\label{fig:cavity-opening-simul}
\end{figure}

Fig.~\ref{fig:cavity-opening-simul} shows the density profiles at
$t=t_\nu/10$ for discs with $\mathcal{R}=10^5$ and four different
values of $R_{\rm inf}$, as they result from numerical
simulations. The plain line corresponds to $R_{\rm inf}=0.65$ au as
before (nominal case). The dotted line corresponds to a four times
smaller value of $R_{\rm inf}$. All other parameters are the same in
the two simulations. In both cases, the Jupiter mass planet has opened a gap
in the disc and migrated down to $r_p\approx 3.85$. The outer disc
profiles almost overlap. The inner disc profiles, however, present a
huge difference\,: the maximum density is nearly twice bigger in the
small $R_{\rm inf}$ case than in the nominal $R_{\rm inf}$ case. In
the nominal case, the inner disc is being depleted and a cavity is
appearing. In the small $R_{\rm inf}$ case, there is only a gap. 
The tendency to the opening of
deeper cavities with larger $R_{\rm inf}$ is confirmed in the last two
simulations (short dashed and long dashed curves, respectively, in 
Fig.~\ref{fig:cavity-opening-simul}). 

In conclusion,  depending on the width and shape of the gap and on the
position of the planet $r_p$ with respect to the inner edge of the
disc $R_{\rm inf}$, the density profile may show a cavity or
not. Indeed, if the inner edge of the gap falls in the flat profile
zone ($r \gtrsim 10\,R_{\rm inf}$), a classical gap is shaped (case
$\mathcal{R}=10^{4}$ in Fig.~\ref{fig:profils} and case $R_{\rm
inf}=0.16$ in Fig.~\ref{fig:cavity-opening-simul}). If it falls
between $0$ and $\sim 4\,R_{\rm inf}$, the density in the inner disc
cannot grow enough before that the inner edge of the gap is reached 
and the planet seems to open a
cavity (case $\mathcal{R}=10^{5.5}$ in Fig.~\ref{fig:profils}). Thus, the
main parameter to determine whether a planet opens a cavity or not is
the ratio between $r_p$ and $R_{\rm inf}$. 
This should be taken into account in the
interpretation of numerical simulations. To perform
realistic simulations, it is necessary to have a realistic value of
$R_{\inf}$. Thus, our code with coupled 2D and 1D grids
\citep{Crida-etal-2007} seems to be a tool of choice as it can handle
arbitrarily small $R_{\rm inf}$.

Our result differs from the one by \citet{Varniere-etal-2006}. They
claimed that the depletion of the inner disc was faster than viscous
because of the negative torque exerted by the planet. However, a
planet in equilibrium in the middle of the gap simply transfers to the
inner disc the torque that it feels from the outer disc\,; otherwise,
the planet would move with respect to the gap. More precisely, in
type~II migration, the planet generally feels a total negative torque,
and migrates inward, together with the disc and the gap\,; thus, the
negative torque that the planet exerts on the inner disc is a bit {\it
smaller} in magnitude than the one it feels from the outer disc. The
difference corresponds to the loss of angular momentum of the
planet. In the absence of the planet, there would be gas in the gap\,;
this gas would also feel a negative torque from the outer disc, exert
a smaller negative torque on the inner disc, and migrate inward losing
angular momentum. If the planet has the same mass as the gas that was
initially in the gap, it has exactly the same effect on the
disc. Consequently, we believe that the presence of the planet does
not modify substantially the evolution of the inner disc --\,except
maybe if the planet is much more massive than the disc, which was
indeed the case in \citet{Varniere-etal-2006} analysis. This is
confirmed in our simulations (see Section~\ref{sec:depletion}).

\section{Modelling the migration rate}

\label{sec:interpretation_migr}

It seems quite logical that, when the planet opens a clean gap, its
migration follows a proper type~II regime.  With a Jupiter mass planet
and a disc aspect ratio of $0.05$, this happens for
$\mathcal{R}>10^{5}$ (see Figs~\ref{fig:profils} and \ref{fig:Migr_tvisq}). 

However, for smaller Reynolds numbers, the gap is not completely
gas-proof.  The gas in the gap has two major consequences\,: (i) it
partially sustains the outer disc, effectively reducing the torque
felt by the planet from the outer disc (ii) it exerts a corotation
torque on to the planet.  The possibility of gas flowing through the gap
decouples the planet from the gas
evolution.

In this Section  we show with a simple model that taking into
account these effects allows us to explain the
evolution of the planet as a function of the various parameters. Our
model is based on previous works on the corotation torque
\citep{Masset2001}, the viscous evolution of accretion discs (LP74),
and the shape of gaps \citep{Crida-etal-2006}.

\subsection{Classical type~II torque}
\label{par:Classical}

In an accretion disc, the viscous stress is such that angular
momentum flows outward while matter falls on to the star. In a
Keplerian, circular disc with $\nu$ and $\Sigma$ independent of the
radius, the torque exerted by the part of the disc extending from a
given radius $r_0$ to infinity on the inner part $\{r<r_0\}$ is $T_\nu
= -3\pi \Sigma\nu{r_0}^2\Omega_0$ (it can be easily found from the
strain tensor). It causes a mass flow of gas $F$, carrying the
equivalent angular momentum\,: $T_\nu=F\,{r_0}^2\Omega_0=(2\pi
r_0\Sigma v_r) {r_0}^2\Omega_0$, where $v_r$ is the radial
velocity of the gas. 
In this model, $v_r=-\frac32\frac{\nu}{r_0}$, which can also
be found from the Navier-Stokes equations. This gives the following
equality, which we will use further\,:
\begin{equation}
\nu = -\frac{2v_r}{3}r_0\ .
\label{eq:nu-v_r}
\end{equation}

If a planet opens a deep gap in such a disc, no gas flow is allowed
through the planetary orbit. The outer disc is maintained outside of the
gap by the planet, and an equilibrium is reached so that the planetary
torque balances $T_\nu$. Consequently, the planet feels from the outer
disc the torque $T_\nu$. This torque is proportional to the viscosity
and not to the planet mass. This is the case of standard type~II
migration.

In a more realistic, viscously evolving disc, the scheme for type~II
migration is the same, but the above formula for $T_\nu$ is no longer
valid. In that case, the equations of LP74 provide the density, the
viscous torque, and the radial velocity as a function of radius and
time. In our case of a disc with $R_{\rm inf}>0$, it gives\,:
\begin{eqnarray}
\label{eq:LP74(25)}
T_\nu & = & 3\pi\nu\Sigma_0T^{-5/4}\left(h-h_{\rm inf}\right)\exp\left(-\frac{ar^2}{T}\right)\\
\label{eq:LP74(25)Sigma}
\Sigma_{\rm LP74} & = & T_\nu / \left(3\pi\nu\sqrt{r}\right)\\
F & = & -\frac{\partial T_\nu}{\partial h}\\
\label{eq:LP74(25)vr}
v_r & = & F/\left(2\pi r\Sigma_{\rm LP74}\right)\ .
\label{rad-vel}
\end{eqnarray}
where $h=r^2\Omega=\sqrt{r}$ is the specific angular momentum. Notice
that equation~\eqref{eq:LP74(25)} is exactly equation~(25) in LP74, while
equation~\eqref{eq:LP74(25)Sigma} is equivalent to equation~\eqref{eq:LBP}.

Thus, in standard Type~II migration, we consider that the planet feels
from the disc a torque
\begin{equation}
T_{\rm II} = Fh = 2\pi r^+\Sigma_{\rm LP74}(r^+)v_r(r^+)\sqrt{r^+},
\label{eq:Hdot_normal}
\end{equation}
where $r^+=r_p+x_s$ is the radius of the external edge of the gap,
and $\Sigma_{\rm LP74}$ and $v_r$ come from equations~\eqref{eq:LP74(25)Sigma} and
\eqref{eq:LP74(25)vr} respectively.

\subsection{Torque exerted on the outer disc by the gas in the gap}

The gas in the gap, the density of which is denoted $\Sigma_{\rm
gap}$, exerts on the outer disc a positive viscous torque $T_{(i)}$
that is given by equation~\eqref{eq:Hdot_normal}, with $\Sigma_{\rm gap}$
instead of $\Sigma_{\rm LP74}$ and the opposite sign. This torque
partially sustains the outer disc, and therefore needs to be
subtracted from the torque that the planet would suffer from the
outer disc if the gap were clean (given by equation~\eqref{eq:Hdot_normal} ).
So, denoting by $f$ the ratio $\Sigma_{\rm gap}/\Sigma_{\rm LP74}$ we
have
\begin{equation}
T_{(i)}=-f T_{\rm II}\ .
\label{T_i}
\end{equation}

\begin{figure}
\includegraphics[width=0.7\columnwidth, angle=270]{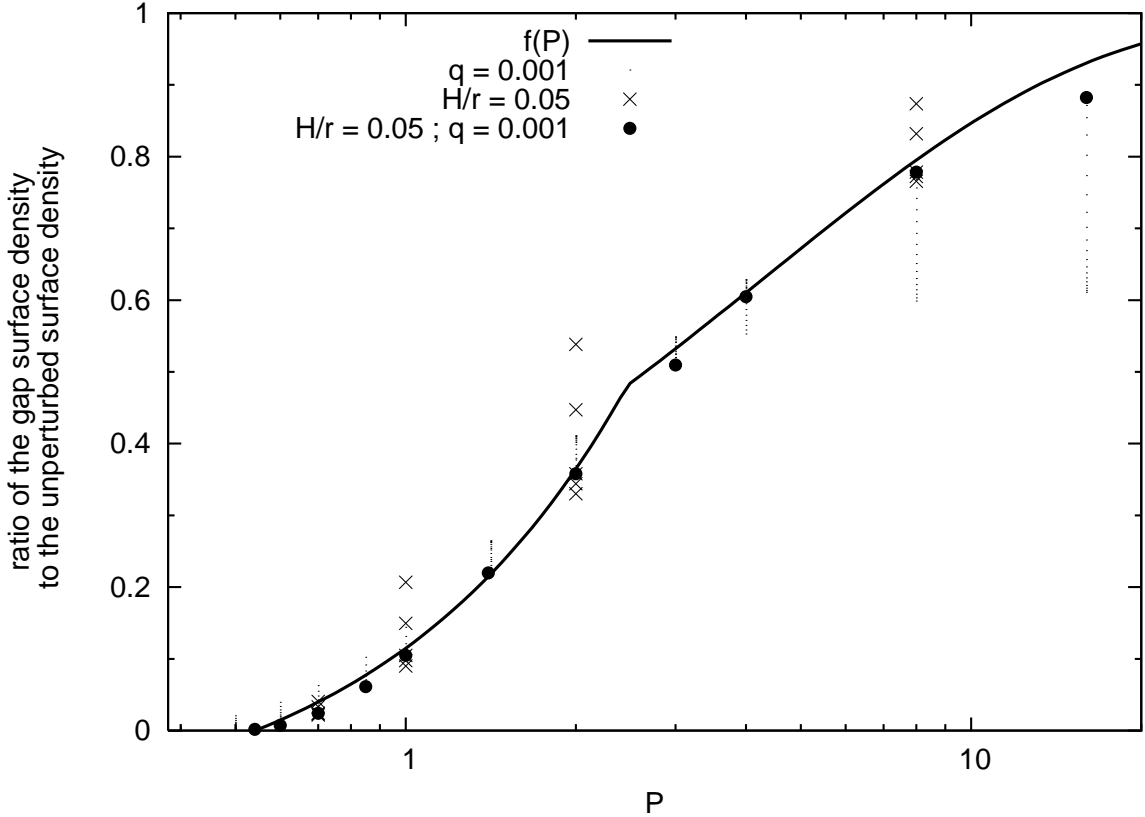}
\caption{Gap depth (measured as ratio of the gap surface density to
the unperturbed density at $r=r_p+2R_H$)
 as a function of $\mathcal{P}$.  The
data points for each value of $\mathcal{P}$ are obtained from the
integration of equation~\eqref{eq:Crida}, assuming
different values of $\nu$ and $H/r$ and keeping $q=10^{-3}$ (points) or
different values of $\nu$ and $q$ and keeping $H/r=0.05$ (crosses)\,; the big
dots correspond to the gap depths obtained for different values of
$\nu$ and keeping both $H/r=0.05$ and $q=10^{-3}$ (see text for a
more precise description of the sets of measures). The bold line is an
approximate fit of the data.}
\label{fig:f(P)} 
\end{figure}

We now discuss how to evaluate $f$ in practice.
We have presented in Section~\ref{sec:interpretation_dp} a way
to compute semi-analytically the gap profile and the gap
depth. However, making a step by step integration until the bottom of
the gap it is not very convenient. Consequently, we looked for a
simple empirical formula for the gap depth as a function of the viscosity, the
aspect ratio of the disc and the planet mass. \citet{Crida-etal-2006} showed
that the density inside the gap is less than $10$ per cent of the
unperturbed value (i.e. $f<0.1$) if and only if\,:
\begin{equation}
\mathcal{P} = \frac34\frac{H}{R_H} + \frac{50}{q\mathcal{R}}
\lesssim 1\ .
\label{eq:P}
\end{equation}
Using equation~\eqref{eq:Crida}, we have computed the depth of the gap for
various values of the parameter $\mathcal{P}$. For each value of
$\mathcal{P}$, we impose $q=10^{-3}$ and $H/r=0.05$, and find the
corresponding viscosity. Then, we use these parameters in
equation~\eqref{eq:Crida}\,; the obtained gap depth is shown as big dots in
Fig.~\ref{fig:f(P)}.  We repeat the same operation for $q$ ranging
from $5\times 10^{-4}$ to $2\times 10^{-3}$\,; the results are
reported as crosses in Fig.~\ref{fig:f(P)}. Furthermore, we impose
$q=10^{-3}$ and $\nu=0$, and find the corresponding $H/r$ and the
resulting gap depth. We repeat the operation for $\nu$ ranging up to
its maximum possible value compatible with $\mathcal{P}$\,; the
results are shown as dots in Fig.~\ref{fig:f(P)}.  Finally, for a
given value of $\mathcal{P}$, different depths are observed because
the shape of the gap is not the same whether the viscosity or the
pressure dominates. However, there is a very clear tendency for deeper
gaps with decreasing $\mathcal{P}$, as could be expected. The bold
line in Fig.~\ref{fig:f(P)} shows the fit that we adopt\,:
\begin{equation}
f(\mathcal{P})=\begin{cases}
(\mathcal{P}-0.541)/4 & \text{ if }\mathcal{P}<2.4646 \\
1-\exp(-\frac{\mathcal{P}^{0.75}}{3}) & \text{ if }\mathcal{P}\geqslant 2.4646
\end{cases}
\label{eq:f(P)}
\end{equation}
where $f(\mathcal{P})$ stands for $\Sigma_{\rm gap}/\Sigma_{\rm LP74}$.

As the gap depth given by equation~\eqref{eq:Crida} is not very precise and
shows variations at fixed $\mathcal{P}$, this fit is clearly a crude
approximation of the real depth. However, the evolution of the gap
depth as a function of the parameters is satisfactorily described by
$f(\mathcal{P})$, in particular if one imposes $q=10^{-3}$ and
$H/r=0.05$, which is the most common case for us. As we are looking
for a simple model, this approximation is sufficient for our purpose.

\subsection{Corotation torque}
\label{corotation_torque}

In \citet{Masset2001}, the corotation torque is computed as
$$ T_C = \frac32{\Omega_p}^2{x_s}^4\Sigma_\infty\,
\mathcal{F}\left(z_s\right)\ ,$$ where $\Sigma_\infty$ is the
unperturbed surface density (assumed to be independent of radius),
$z_s=(\mathcal{R}/2\pi)^{1/3} x_s/r_p$, the function
$\mathcal{F}(z)=z^{-3} - z^{-4}g(z)/g'(z)$ and $g$ is a linear
combination of the Airy functions Ai and Bi defined in equation~(A18)
of \citet{Masset2001}. This formula can easily be generalised for an
arbitrary profile of $\Sigma$ \citep[see ][]{Ward1991}, and be rewritten as:
$$T_C = \frac34{\Omega_p}^2{x_s}^4\Sigma\,
\frac{{\rm d}\log (\Sigma/B)}{{\rm d}\log r}\,
\hat{\mathcal{F}}\left(\frac{1}{(x_s/r_p)^{3}\mathcal{R}}\right)$$
where $B$ is the second Oort constant ($B=\Omega/4$ in Keplerian rotation), 
and $\hat{\mathcal{F}}(z)=4\mathcal{F}\left((z/2\pi)^(1/3)\right)$.

The function $\hat{\mathcal{F}}(z)$ is shown in graphical form in Fig.~2
in \citet{Masset2001}. For $z<0.1$ (which is our case  
if we consider $x_s = 2R_H$, $q=10^{-3}$ and $\mathcal{R}>10^{3.6}$)
one has $\hat{\mathcal{F}}(z)\approx 20z$. Thus, we may write\,:
$$T_C = \frac34\,20x_s\Omega_pr_p\Sigma\nu\frac{{\rm d}\log (\Sigma/B)}{{\rm d}\log r}$$

In \citet{Masset2001}, the approximation that $\nu$ and $\Sigma$ are
independent on $r$ is made. Thus, we can use equation~\eqref{eq:nu-v_r} and
write equivalently\,:
\begin{equation}
T_C = -10\,x_s\Omega_p{r_p}^2\Sigma v_r\,\frac{{\rm d}\log (\Sigma/B)}{{\rm d}\log r}
\label{eq:T_C}
\end{equation}
Here, $\Sigma$ is the density inside the gap, which is equal to
$\Sigma_{\rm LP74}(r^+)f(\mathcal{P})$, and $v_r$ is the radial
velocity at $r=r_p$, which is given in
equation~\eqref{eq:LP74(25)vr}. The term $\frac{{\rm d}\log
(\Sigma/B)}{{\rm d}\log r}$ is computed from the unperturbed density
$\Sigma_{\rm LP74}(r)$, assuming a Keplerian rotation of the disc.

\subsection{Total torque exerted on the planet}

The total torque felt by the planet is therefore , $T_p=T_{\rm
II}-T_{(i)}+T_C$. From equations~\eqref{eq:Hdot_normal}, \eqref{T_i}
and \eqref{eq:T_C}, the total torque reads\,:
\begin{equation}
\begin{array}{rcl}
T_p & = & 2\pi r^+\Sigma_{\rm LP74}(r^+)v_r(r^+)\sqrt{r^+}\times\\
 & &
 \left[1-f(\mathcal{P})-\frac{15}{2\pi}\frac{x_s}{r^+}\frac{\Omega_p{r_p}^2}{\sqrt{r^+}}\frac{v_r(r_p)}{v_r(r^+)}
 f(\mathcal{P}) \frac{{\rm d}\log (\Sigma/B)}{{\rm d}\log r} \right]
\end{array}
\label{eq:T_p}
\end{equation}
with $\mathcal{P}$ and $f$ defined respectively in
equations~\eqref{eq:P} and \eqref{eq:f(P)}, while $r^+=r_p+x_s$,
$x_s=2R_H$, and $\Sigma_{\rm LP74}$ and $v_r$ come from
equations~\eqref{eq:LP74(25)}-\eqref{eq:LP74(25)vr}.  This expression
involves (directly or via $v_r$ or $\Sigma_{\rm LP74}$) the viscosity
of the disc $\nu$, its aspect ratio $H/r$, the planet to primary mass
ratio $q$, the radius of the planetary orbit $r_p$, and the radius of
the inner edge of the disc $R_{\rm inf}$. In the following, we test it
against numerical simulations for a wide range of these parameters.

\subsubsection{Dependence on the Reynolds number}
\begin{figure}
\includegraphics[width=0.875\columnwidth, angle=270]{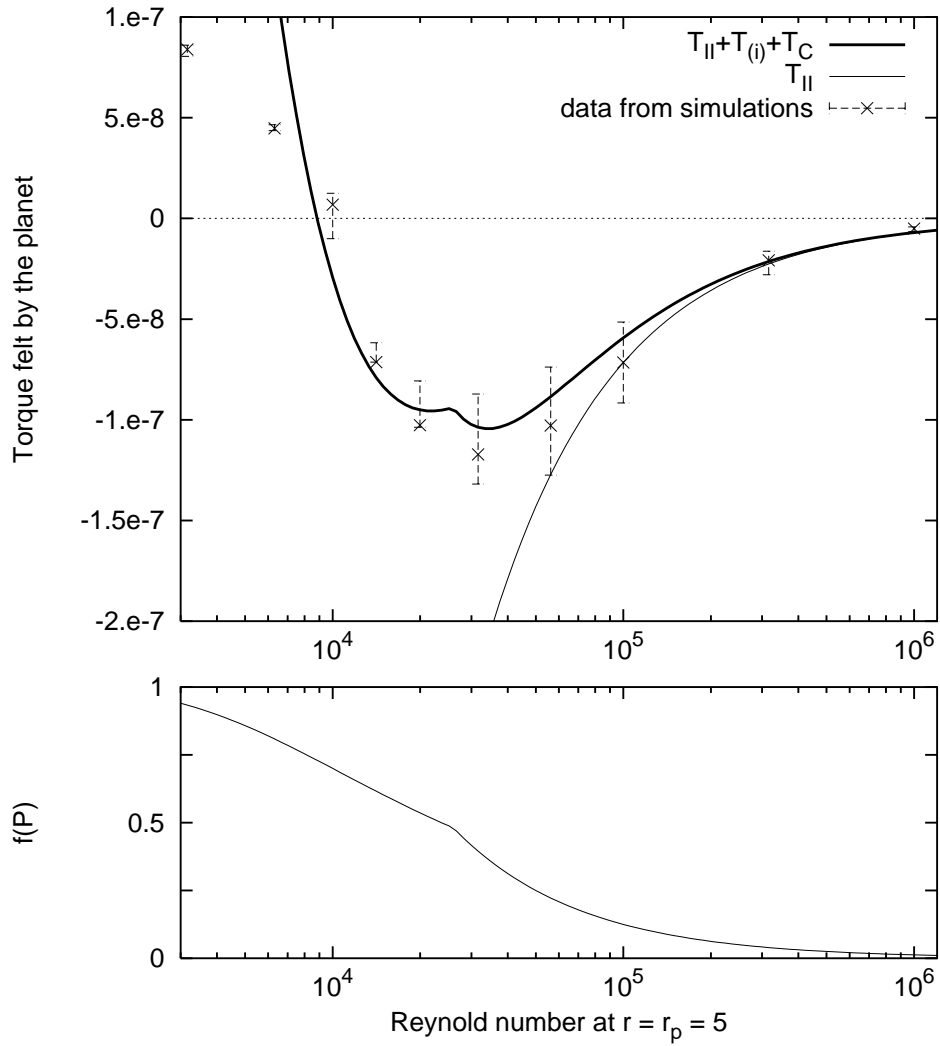}
\caption{Top panel\,: torque felt by the planet, expressed in the
units defined in Section~\ref{sec:units}, as a function of the Reynolds
number. Bottom panel\,: the gap depth $f(\mathcal{P})$ as a function
of the same parameter.}
\label{fig:model-test}
\end{figure}

In the top panel of Fig.~\ref{fig:model-test} we plot the total torque
felt by a Jupiter mass planet located at $r_p=5$ au in a disc
with $0.05$ aspect ratio and $R_{\rm inf}=0.65$ au as a function
of $\mathcal{R}$. The torque given by equation~\eqref{eq:T_p} with $t =
t_\nu/100$ is plotted as a bold line.  The crosses with the error bars
represent the torque felt by the planet in the numerical simulation,
measured at $t=t_\nu/100$ (so that the planet is still at about
$r_p=5$ au). The error bars correspond to the maximum and minimum
migration rates measured between $t_\nu/200$ and $t_\nu/50$. The
torque $T_{\rm II}$, corresponding to classical type~II migration, is
drawn as a thin line\,; it is proportional to
$\nu\propto\mathcal{R}^{-1}$. The bottom panel shows, as reference,
the gap depth $f(\mathcal{P})$ as a function of $\mathcal{R}$.

As one can see in Fig.~\ref{fig:model-test}, the model reproduces
very well the general tendency. For $\mathcal{R}>10^5$,
$f(\mathcal{P})<0.125$ and $T_p$ is close to $T_{\rm II}$, negative
and proportional to the viscosity, as expected in proper type~II
migration\,; in other words, $T_{\rm II}$ dominates $T_{(i)}$ and 
$T_C$ in the calculation of $T_p$. For lower Reynolds numbers, the
effects of the gas in the gap counterbalance the classical $T_{\rm
II}$. For $\mathcal{R}<10^{4}$, $f(\mathcal{P})>0.7$ and the total
torque is positive. Both the measures from the simulations and the results of
equation~\eqref{eq:T_p} show a minimum of the torque felt by the planet at
about $\mathcal{R}=10^{4.5}$. This shows that type~II migration speed
is bounded. One could think that an increase in the gas viscosity, and
therefore of the gas accretion rate, would increase the type~II migration
speed of a giant planet. In reality, this favours the filling of the
gap, which decouples the planet from the disc
evolution.

Notice however that for very low Reynolds
number (high viscosity) our model reaches its validity limits. 
In fact, our model predicts a very fast outward migration, which is
not observed in the simulations. In these cases, 
$f(\mathcal{P})$ is more than $0.8$ (essentially no gap), 
so that the basic idea on which our model is built is no longer
valid. Eventually, for large enough viscosity, the planet has to
migrate inward, at a Type~I migration rate.

\subsubsection{Dependence on the planet mass}
\begin{figure}
\includegraphics[width=0.7\columnwidth, angle=270]{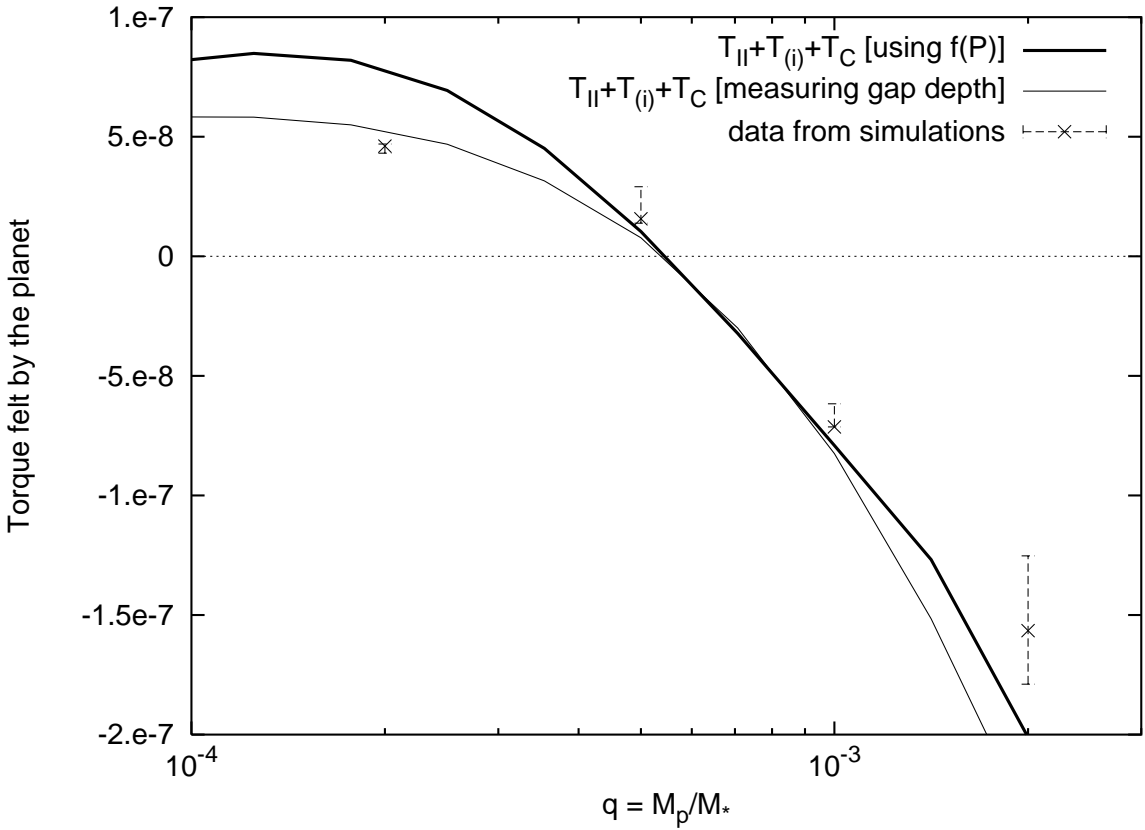}
\caption{Torque felt by a planet as a function of its mass.}
\label{fig:model-Mp}
\end{figure}

In Fig.~\ref{fig:model-Mp}, the torque exerted on a planet placed at
$r_p=5$ in a disc with $\mathcal{R}=10^{4.15}$, $H/r=0.05$ and $R_{\rm
inf}=0.65$ is plotted as a function of the planet mass. The bold line
corresponds to our model, and the crosses correspond to numerical
experiments (with the error bars computed with the same prescription
as before). Once again, we recover the observed trend.  The
discrepancy at very low mass comes in fact from the estimation of the
gap depth $f(P)$\,; for $q<5\times 10^{-4}$, the density in the `gap'
is larger than $0.8$, so that the gap is shallow and $f$ is not very
accurate. Using directly equation~\eqref{eq:Crida} instead of prescription
equation~\eqref{eq:f(P)} for the gap depth gives a better result, shown as a
thin line\,; this validates again our model.

The torque felt by the planet is a decreasing function of its
mass. The explanation for this behaviour is quite easy\,: the lighter
is the planet, the shallower is the gap, and the more important are
the effects of the gas in the gap on the total torque.
Symmetrically, the more massive is the planet, the deeper is the gap
it opens, and thus the closer to type~II migration is its migration
regime.

This result is relevant for extra-solar planets as it can explain why, 
of the many couples of resonant planets, typically the most massive object
is the outermost one. Indeed, resonant capture is possible only if 
the outer planet migrates inward faster than the inner planet which,
according to Fig.~\ref{fig:model-Mp},  requires that the outer planet
is the most massive.

\subsubsection{Dependence on planet accretion}

The planet mass is not necessarily constant. Planets are supposed to
accrete gas continuously. Gas accretion by the planet perturbs the gas
flow through the planetary orbit\,; in addition, mass accretion exerts
an additional torque on the
planet \citep[][]{LubowDAngelo2006}. Therefore, in principle, accretion
could change the dynamics. We have implemented planetary accretion
following the recipe of \citet{Kley1999}\,: it consists in removing a
fraction of the material in the Hill sphere of the planet and adding
it to the mass of the planet. The accretion rate, expressed as a
fraction of mass removed per time unit is imposed as an input
parameter. More precisely, we apply the input removal rate $f_{\rm
accr}$ in the inner part of the Hill sphere (extended up to
$0.45\,R_H$), and two thirds of $f_{\rm accr}$ in the region from
$0.45\,R_H$ to $0.75\,R_H$.

\begin{figure}
\centering
\includegraphics[width=\columnwidth, angle=270]{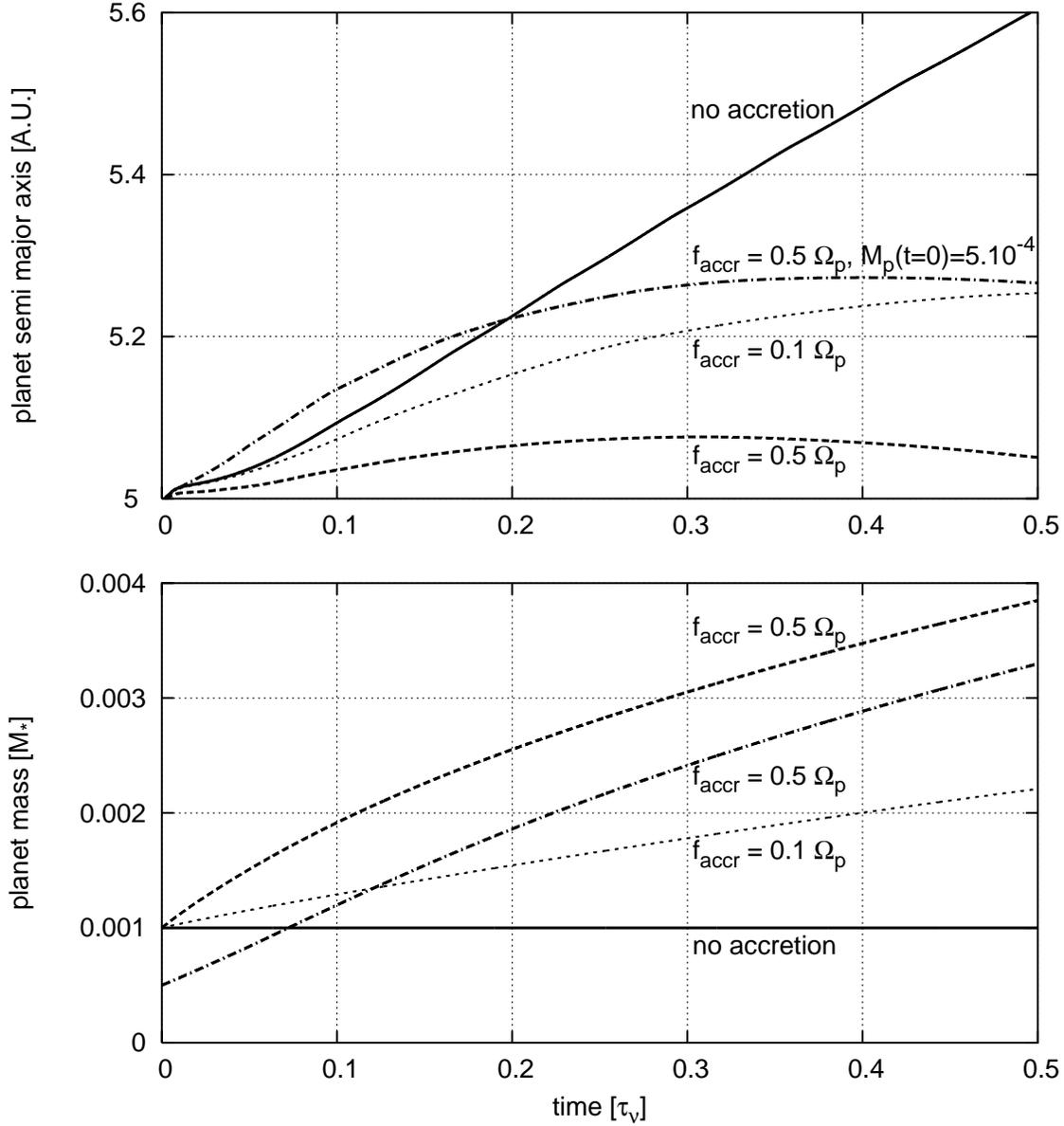}
\caption{Top panel\,: migration of accreting giant planets in a disc
with $\mathcal{R}=10^{3.8}$ and $H/r=0.05$. Bottom panel\,:
corresponding evolution of their masses.}
\label{fig:migr-accretion}
\end{figure}

Considering $\mathcal{R}=10^{3.8}$ and $H/r=0.05$, we performed
simulations of a Jupiter mass planet initially on a circular orbit at
5 au, with input removal rates $f_{\rm accr}=0.5\Omega_p$ and
$f_{\rm accr}=0.1\Omega_p$. The results in terms of migration
and accretion are shown in Fig.~\ref{fig:migr-accretion} and compared
with the no accretion case. Another simulation with input removal
rate $0.5\Omega_p$ and initial mass of the planet $5\times
10^{-4}$ is also performed. The results show that accretion does not
prevent outwards migration. The comparison between the top panel
(migration) and the bottom panel (mass evolution) shows that the
migration speed seems to be governed by the planet mass, and not by
the accretion rate. While the accretion rate is constant with time in
all cases, the migration rate is not, and its evolution follows the
evolution of the planet mass. As expected from the discussion above,
the outwards migration rate is slower for more massive planets\,; if
the planet reaches 3 Jupiter masses, its migration vanishes\,; the
migration is directed inwards for more massive planets.

\subsubsection{Dependence on the radii of the planetary orbit and of the disc inner edge.}
\label{subsec:rp/rinf}

One may wonder about the dependence of the migration rate on the
planet's location in the disc. This is governed by the radial
dependence of viscosity and aspect ratio.

If the viscosity is constant, the Reynolds number $\mathcal{R}$
increases as $\sqrt{r}$. Thus, the depth of the gap opened by the
planet {\it increases} with the planet's location $r_p$. As a
consequence a planet should migrate inward if it is sufficiently far
from the star, and outward if it is sufficiently close. The migration
paths therefore converge towards the stationary solution, at some
specific radius, dependent on viscosity and planet mass.
 
However, if the viscosity of the disc depends on radius as described
in an $\alpha$ model \citep{ShakuraSunyaev1973}, the Reynolds number
is independent of radius.  If the disc is flared, $H/r$ increases with
$r$ and therefore the depth of a gap opened by the planet {\it
decreases} with $r_p$. Consequently the planet tends to move outward
if it is far from the star and inward if it is close. Any stationary
solution would therefore be unstable.

\begin{figure}
\includegraphics[width=0.7\columnwidth,angle=270]{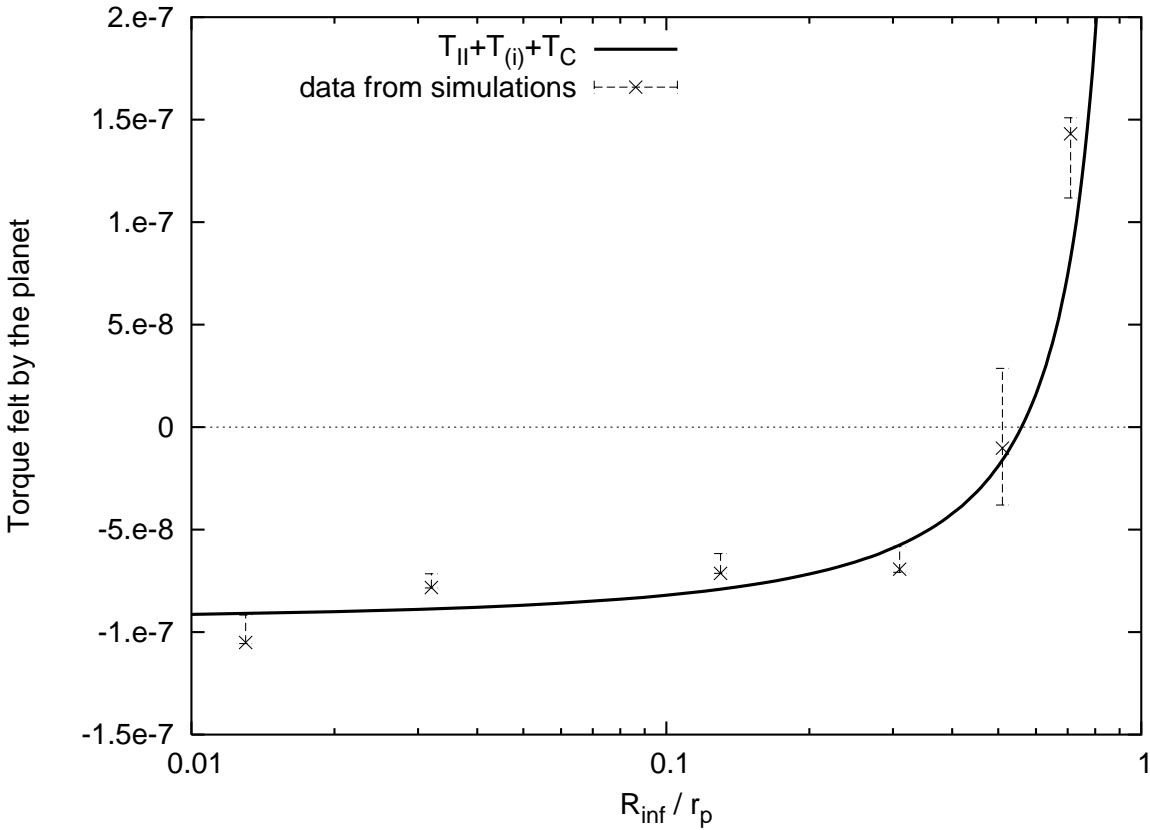}
\caption{Torque felt by a Jupiter mass planet at $r_p=5$ with
$\mathcal{R}=10^{4.15}$, $H/r=0.05$, at $t=t_\nu/100$, as a
function of the radius of the inner edge of the disc $R_{\rm inf}$.}
\label{fig:model-rinf}
\end{figure}

If the disc is not flared and the Reynolds number does not depend on
$r$, then the behaviour of the planet depends on the ratio $r_p/R_{\rm
inf}$.  This effect is subtle.  If the ratio $r_p/R_{\rm inf}$ is
smaller, the planet is in a location of the disc where $\Sigma$ and
$v_r$ have a steeper positive slope, and $v_r$ is also larger in
absolute value, following equation~\eqref{eq:LP74(25)vr}. In our model, the
gas radial velocity multiplies all the torques that appear in
equation~\eqref{eq:T_p}. Therefore, the only dependence on $r_p/R_{\rm
inf}$ is through the gradient of the gas radial velocity
$v_r(r_p)/v_r(r^+)$ and the gradient of the logarithm of the density
that appear in equation~\eqref{eq:T_p} for the corotation
torque. Fig.~\ref{fig:model-rinf} shows as a bold line the torque
felt by a Jupiter mass planet for $\mathcal{R}=10^{4.15}$ at
$t=t_\nu/100$, as expected in our model. The gradient effect becomes
important when $R_{\rm inf}/r_p\geqslant 1/2$. The crosses with the
error bars correspond to measures from numerical simulations. Once
again, our model reproduces very well the observed trend.

This result might be relevant to explain the existence of hot
or warm Jupiters. Although several solutions have been proposed,
the issue of why these planets did not migrate all the way into their
parent stars remains open. Figure~\ref{fig:model-rinf} shows that,
for the parameters used in that calculation, a planet
approaching the inner edge of the disk has eventually to stop at the
location where the migration rate turns from negative to
positive. Therefore, the survival of some hot\,/\,warm Jupiters
against migration seems to be a natural consequence of the local
surface density gradient at the inner edge of the disk, predicted by
LP74. Notice however that some hot Jupiters are so close to the
central star that they are presumably well within the corotation
radius. As the disk inner edge should not have been closer than the
corotation radius, these planets somehow managed to migrate past the
inner edge of the disk. In the framework of our model this may be
possible for massive planets in low viscosity disks, opening deep and
clean gaps. In these cases, the planets presumably stopped migrating
when their outer 1:2 mean motion resonances reached the inner disk
edge, as proposed in \citet{KuchnerLecar2002}. In conclusion, the
dynamical diversity of the exoplanets could be a consequence of the
physical diversity of the protoplanetary discs.

\section{Effect of planet migration on the inner disc depletion}

\label{sec:depletion}

In Fig.~\ref{fig:model} we have shown that in the case of a non-migrating
planet, the density profile in the inner part of the disc coincides with the
unperturbed profile given by LP74 equations. This showed
that the overall mass flow through the disc was unperturbed by the presence of
the planet.

\begin{figure}
\includegraphics[width=0.7\columnwidth, angle=270]{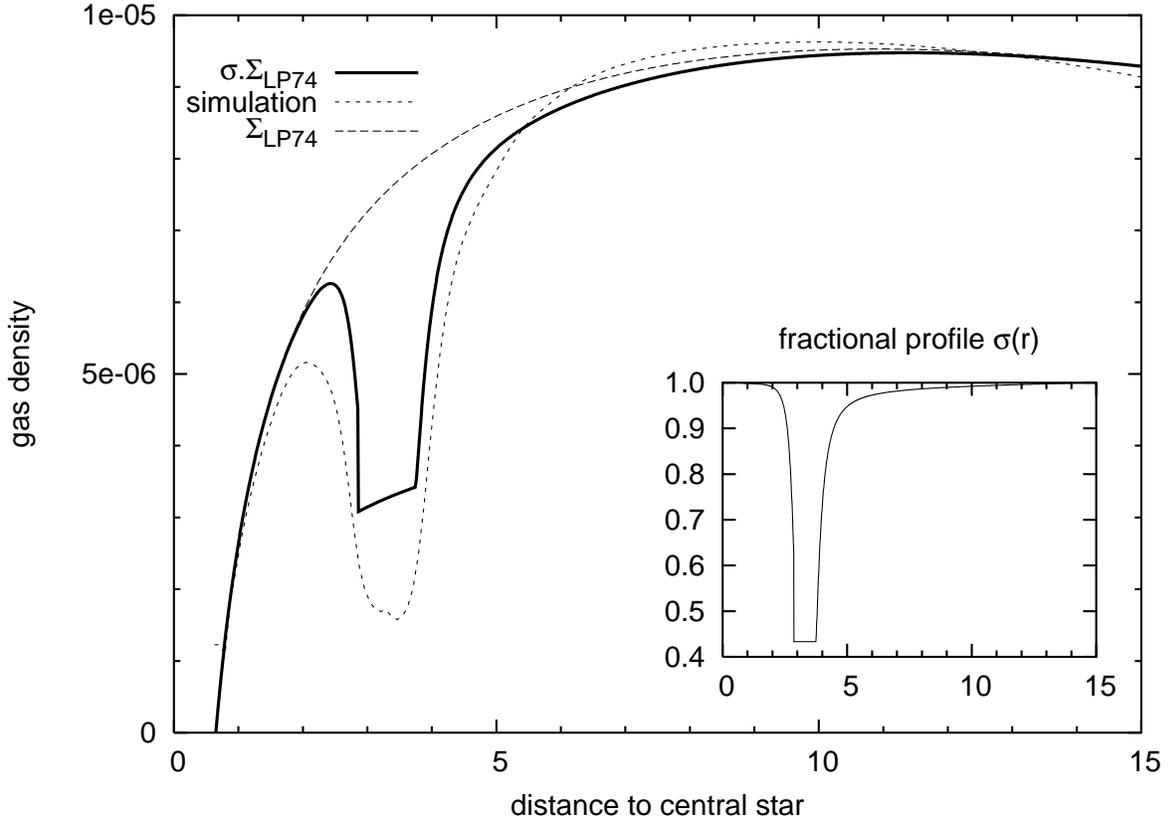}
\caption{Gap profile in an evolving disc with
$\mathcal{R}=10^{4.5}$\,: the thin line results from numerical
simulation while the bold line comes from our simple model. Dashed
line\,: the unperturbed profile $\Sigma_{\rm LP74}(r,t_\nu/2)$ from
equation~\eqref{eq:LBP}. Secondary panel\,: `fraction profile'
$\sigma(r)$ of the gap integrated from equation~(14) of \citet{Crida-etal-2006}.}
\label{fig:mass-def}
\end{figure}

Fig.~\ref{fig:mass-def} is the same as Fig.~\ref{fig:model} but for a case
where the planet moves inward at slow rate. In this case there appears to be a
deficit in the inner part of the disc, while the outer part shows a little
over-density, relative to the unperturbed case and to the prediction
of our simple model. This reveals that only a
fraction of the unperturbed mass flow effectively passes from the outer to the
inner disc.

Based on the model developed above, we can illustrate this fact and
develop an {\it a priori} estimate of the inner
disc mass deficit.

Due to the migration of the planet, the surface of the inner disc
shrinks at a speed\,:
$$
\dot{A}_p= 2\pi r_p \dot{r_p}\ .
$$
In the unperturbed case, if we draw a fictitious boundary between the inner and
the outer disc moving at the radial velocity of the gas, the surface of the
inner disc shrinks at a speed\,:
$$
\dot{A}_{\rm unp}= 2\pi r_p v_r\ .
$$
Thus, for the inner mass of the disc to remain the same as in the unperturbed
case, the mass flow into the inner disc through a planet's gap must be equal
to\,:
$$
\dot M_0 = -(\dot{A}_{\rm unp}-\dot{A}_p)\Sigma = 2\pi
r_p(\dot{r_p}-v_r)\Sigma
$$
where $\Sigma$ is the unperturbed gas density, to be identified with
$\Sigma_{\rm LP74}$ above.
The mass flow into the inner disc through the planet's gap corresponds to the
angular momentum loss of the outer disc. It can be expressed as\,:
\begin{equation}
\dot M'=-\dot{J_{\rm o}}/j
\label{dotMprime}
\end{equation}
where $J_{\rm o}$ is the total angular momentum of the outer disc and $j$ is
the specific angular momentum. Obviously, $\dot{J_{\rm o}} =
\dot{J_{\rm o}}_{\rm ,unp}-\dot{J_p}$, where $\dot{J_{\rm o}}_{\rm ,unp}$ is
the change of angular momentum of the outer disc in the unperturbed case and
$\dot{J_p}$ is the change of angular momentum of the planet.

Thus, the mass flow deficit into the inner disc is\,:
\begin{equation}
\delta \dot M = \dot M_0 -\dot M'= 2\pi r_p(\dot{r_p}-v_r)\Sigma + (\dot{J_{\rm
o}}_{\rm ,unp}-\dot{J_p})/j\ .
\label{massflux}
\end{equation}
Now, remembering that $\dot{J_p}=\dot{r_p}M_p/2\sqrt{r_p}$ and that
$\dot{J_{\rm o}}_{\rm ,unp}=2\pi r_p v_r\Sigma j$, the expression above
becomes\,:
\begin{equation}
\delta \dot M = \frac{\dot{J_p}}{j}\left(4\frac{\mu}{q}-1\right)\ ,
\label{eq:deltadotM}
\end{equation}
where $\mu=\pi r^2 \Sigma/M_*$ and $q=M_p/M_*$ are the reduced masses
of the inner disc and the planet.

Some remarks on equation~\eqref{eq:deltadotM} are in order. If the planet
does not migrate, $\dot{J_p}=0$ and thus there is no mass flow deficit
into the inner disc, in agreement with Fig.~\ref{fig:model}. If
$q=4\mu$ there is also no mass deficit in the inner disc. This case
corresponds to $\dot{r_p}=v_r$, that is to a planet that migrates
inward at the same speed of the unperturbed gas. We recall that this
is the maximal migration speed of a planet in type~II migration. So,
if $q<4\mu$, equation~\eqref{eq:deltadotM} does not apply.

Finally, to compute the mass of the inner part of the disc (and therefore the
cavity depth) one can follow the approach illustrated in
Section~\ref{sec:interpretation_dp} and multiply the result by\,:
\begin{equation}
1-\delta\dot M/\dot{M}_0=1-\cfrac{q/4\mu-1}
{ \cfrac{q}{4\mu}\cfrac{ \dot{J_{\rm o}}_{\rm ,unp} }{\dot{J_p}} - 1 }\ .
\label{eq:deltadotMrelat}
\end{equation}

\vspace{12pt}

From equation~\eqref{eq:deltadotMrelat}, it also appears that if the planet
mass is large with respect to the disc mass, then the mass flow
deficit $\delta\dot M/\dot{M}_0$ tends to $\dot{J_p}/\dot{J_{\rm
o}}_{\rm ,unp}$. If this planet opens a clean gap,
$\dot{J_p}\approx\dot{J_{\rm o}}_{\rm ,unp}$ and $\dot{r_p}/v_r\ll 1$,
so that $\delta \dot M \approx \dot{M}_0$. In summary, the two
conditions $q/\mu\gg 1$ and formation of a clean gap are required to
deplete strongly the inner disc and form a deep cavity.

\begin{figure}
\includegraphics[width=0.7\columnwidth, angle=270]{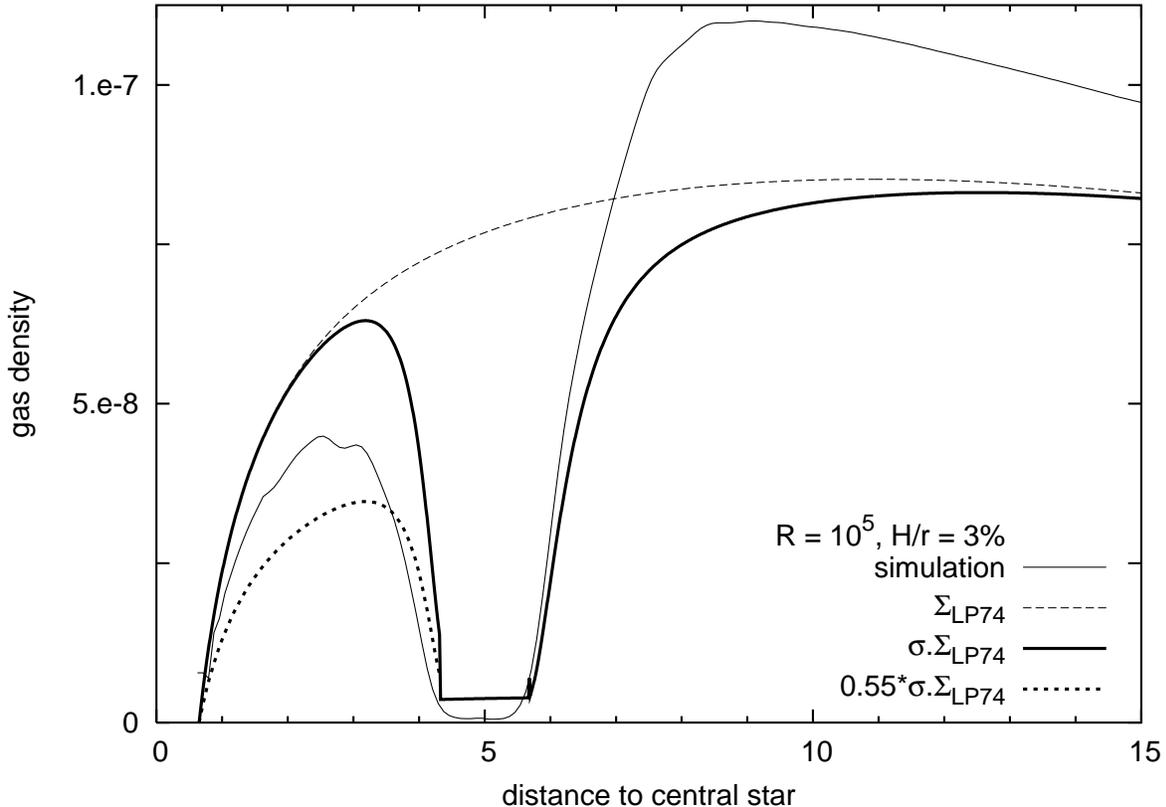}
\caption{Gap profile in a small mass evolving disc with
$\mathcal{R}=10^{5}$ and $H/r=0.03$\,: the thin solid curve results from
a numerical simulation while the bold line comes from our simple
model. Thin dotted curve\,: the unperturbed profile $\Sigma_{\rm
LP74}(r,t_\nu/2)$ from equation~\eqref{eq:LBP}. Bold dotted curve\,: the
inner disc profile in our model, multiplied by the estimated relative
mass deficit.}
\label{fig:thindisc}
\end{figure}

To confirm this result, we performed a simulation with the same
Jupiter mass planet initially placed on a circular orbit at $r_p=5
au$, and the same initial disc profile as before, except that the gas
density is divided by 100. The Reynolds number is taken equal to $10^5$
and the aspect ratio is decreased to $0.03$ for the planet to open a
cleaner gap than before. Fig.~\ref{fig:thindisc} displays the
resulting disc profile at $t = t_\nu/2$ (thin solid curve)\,: one can
see that the planet did almost not migrate, but the inner disc has
been significantly depleted with respect to the unperturbed profile
(thin dotted curve) and to the profile given by 
equation~\eqref{eq:gap-model} (bold solid curve). As we have seen above,
for `reasonable' values of disc height and viscosity a Jupiter mass
planet does not open a very clean gap. So, in practice, an effective
cavity opening planet would need to have 3--5 Jupiter masses, embedded
in a sub-Jovian mass disc.

The migration rate of the planet at $t = t_\nu/2$ in the simulation is
$ \dot{J_p} = 2.88\times10^{-11}$. The unperturbed density is
$8.48\times 10^{-8}\ M_\odot.au^{-2}$ so that
$q/4\mu=36.84$, and the unperturbed radial velocity of the gas is
$-1.03\times 10^{-5}$. Following
equation~\eqref{eq:deltadotMrelat}, this gives a relative mass deficit in
the inner disc\,:
$$\delta\dot M/\dot{M}_0 = 0.45\ .$$ Our model density profile,
multiplied by $0.55$ gives for the inner disc the bold dotted curve in
Fig.~\ref{fig:thindisc}. This agreement is quite satisfactory,
although --\,admittedly\,-- a better agreement with the real inner disc
profile would be given by a coefficient of about $0.75$.

\section{Gas cavities versus dust cavities}

\label{sec:dust}

Up to this point, we have discussed only the gas distribution of the disc. 
However, the observations reviewed in Section~\ref{sec:observations}
constrain the absence of dust in the inner part of the disc, rather
than the absence of gas. 

The dust distribution and the gas distribution are not necessarily the
same. As recently pointed out by \citet{Rice-etal-2006} \citep[see
also ][]{PaardekooperMellema2006_dust}, the opening of a gap in the
gas disc can act as a filter on the dust. Only the very small dust can
flow with the gas through the gap and refill the inner part of the
disc. Larger dust particles are repelled by the positive gas pressure
gradient at the outer edge of the gap, and consequently they cannot
pass through the planet's orbit.

Our model provides all the ingredients to compute what is the maximal
size of dust particles that can flow with the gas through the gap as a
function of the various parameters of the problem (planet mass,
$\mathcal P$ etc.), as outlined below.

The gas orbital velocity $v_g$ differs from the Keplerian velocity
$v_K$ due to pressure effects as:
$$ \frac{v_g^{\,2}}{r}=\frac{v_K^{\,2}}{r}+
\frac{1}{\Sigma}\frac{{\rm d}P}{{\rm d}r}\ ,$$
where $P$ is the pressure and $\Sigma$ is the surface density.
Using our equation of state (see Section~\ref{sec:units}) and assuming
that sound speed $c_s$ does not change significantly along the gap's
edge, the last term of the above equation can be rewritten as
$(1/\Sigma) {\rm d}\Sigma/{\rm d}r$.  Its maximum value can be obtained
from equation~\eqref{eq:Crida} at $r=r_p+2R_H$.

Once $v_g$ is computed, the outward drift speed $v_r^{(d)}$ of a dust
particle of a given size and density can be computed following
\citet{Weidenschilling77}. The quantity $v_r^{(d)}$ is actually a
speed relative to the gas radial motion. Thus, if the inward radial
velocity of the gas $v^{(g)}_r$ is larger in absolute value than
$v_r^{(d)}$, the dust flows inwards, passing through the planet's
gap. In the opposite case the dust is repelled by the pressure
gradient and accumulates at the edge of the gap. The radial velocity
of the gas at the base of the gap, where the pressure effect on the
dust is maximal, can be computed as
\begin{equation}
v^{(g)}_r={\frac{\dot M'}{2\pi r_p\Sigma f(\mathcal{P})}}
\end{equation}
where $\dot M'$ is given in equation~\eqref{dotMprime}. Assuming that the mass
flux from the outer to the inner disc is not strongly 
perturbed by the presence of the planet (i.e. 
$\dot M' \sim \dot M_0$, which is strictly true in the case of a 
non-migrating planet) one has\,:
\begin{equation}
v^{(g)}_r = \frac{v_r}{f(\mathcal{P})}
\end{equation}
where $v_r$ is given in equation~\eqref{rad-vel}.
Finally, because $v_r^{(d)}$ depends on the dust particle size, the
equation $|v_r^{(d)}|=|v_r^{(g)}|$ gives the maximal size of the particles
that flow in the inner disc with the gas.

However, we think that it is not evident that a real dust cavity
is opened inside the orbit of the planet, even for dust sizes that
cannot pass through the gap.  The dust in the inner disc should
accumulate at the place of the relative maximum of the gas density
distribution, rather than follow the gas in its accretional motion on
to the star.  In fact, the positive density gradient at the inner edge
of the disc is comparable to that at the outer edge of the gap opened
by the planet (see Figs~\ref{fig:profils}
and~\ref{fig:cavity-opening-simul}). So, if the dust is repelled by
the gap edge, it should also be repelled by the inner disc edge. In
this situation a ring of dust should form in the inner disc, its width
depending on the local velocity dispersion due to the
turbulence. Hence, the planet appears opening a wide and deep gap in
the dust distribution \citep{PaardekooperMellema2006_dust}, rather
than a cavity. This, however, may still be consistent with the
observations\,: if these rings of large dust grains are very
narrow (as they may well be) the particles may not contribute
significantly to the observed SED\,; so, the SED observations would
still suggest an inner cavity with small grains (producing a 10 $\mu
m$ feature) and a population extending to larger sizes in the outer
disc.

Possibilities to get rid of the larger dust grains in the
inner disc could be (i) the collisional comminution into smaller
particles that are coupled to the gas and therefore -- as the gas -- 
accrete on to the star, (ii) formation of larger particles or
planetesimals, favoured by the high dust density in the ring and (iii)
sublimation of the dust if the ring is located sufficiently close to
the star. Obviously the situation is complicated. The study of the
dust distribution requires the use of an adapted bi-fluid code,
including irradiation effects. This should be the object of future work.

\section{Conclusion}

\label{sec:conclu}

In this paper, we have performed numerical simulations of giant planets
embedded in a gas disc of equivalent mass. The disc in the simulations
is viscously evolving, reproducing (in absence of planetary
perturbation) the evolution described in LP74. This is made
possible by the use of a new numerical scheme that complements the
usual 2D grid with a 1D grid
spanning over the assumed physical extension of the disc
\citep{Crida-etal-2007}, so that  the global evolution of the disc is
taken into account. This is a key point for the accurate simulation of
both the cavity opening process and of planet migration. 
Indeed, the first phenomenon
depends on the accretion of the inner disc on to the central star\,; 
thus the simulation of the disc evolution down to its
physical inner edge is necessary for reliable results. As for the
planetary migration, proper type~II migration corresponds to the case
where the planet 
follows the disc evolution\,; the global evolution of the disc is thus the
key phenomenon that needs to be reproduced. 

If the disc mass is not negligible with respect to the planet mass, we
found that the planet does not modify substantially the disc profile
with respect to the free viscous evolution\,; it simply opens a more
or less deep gap in the unperturbed profile. Consequently, the opening
of a cavity mainly depends on the position of the
planetary orbit with respect to the inner edge of the disc $R_{\rm
inf}$. Indeed, at this inner edge, the density is $0$. It grows with
$r$ until the inner edge of the gap. Thus, if the gap is close to
the inner edge of the disc, the density in the inner disc does not
reach the maximum value and a cavity appears. As a result, it is much
easier to open cavities is discs with large $R_{\rm inf}$ than in
discs with small $R_{\rm inf}$. The size of this inner radius may vary
in astrophysical discs, depending on the process that governs the fall
of the gas on to the central star. We stress that one should be aware
of the primordial influence of this parameter on the inner disc
evolution when computing numerical simulations.

In the intermediate regime between type~I and type~II migration
--\,where the planet opens a non-gas-proof gap\,-- the migration may
be stopped or reversed. In fact, if the gap is not clean, the gas in
the gap exerts a positive viscous torque on the outer disc, so that
the outer disc doesn't push the planet inward as efficiently as when
the gap is clear. In addition, the corotation torque is positive and
proportional to the density in the corotating zone. Thus, the total
torque felt by the planet may be zero, or even positive, if the
Reynolds number is low enough (Fig.~\ref{fig:model-test}).  We built a
model based on simple, qualitative ideas, that leads to a simple
expression of the total torque felt by a planet in an evolving disc
(equation~\eqref{eq:T_p}~). This enabled us to reproduce the dependence of
planet migration on the various parameters. Viscosity plays a major
role\,: standard type~II migration occurs only when the planet opens a
very clean gap, that is at low viscosity\,; if the viscosity
increases, the gap becomes less depleted and less gas-proof, which
leads to the decoupling of the migration with respect to the disc
evolution.  The role of the planet mass is very intuitive\,: the more
massive is the planet, the deeper is its gap and thus the closer to
standard type~II migration is its behaviour. Last, the role of the
radius of inner edge of the disc $R_{\rm inf}$ is significant only
when it is more than about half the orbital radius of the planet\,: in
that case, the density gradient of the disc at the planet location
increases with $R_{\rm inf}$, which enhances the corotation torque and
makes the planet migrate outward.

In conclusion, under some conditions on the disc parameters, type~II
migration may be avoided for a Jupiter mass planet at $5$ au in an
accreting disc, provided it does not open a very deep gap. This could
explain why all the known giant planets (in the solar system and in
extra-solar systems) are not hot Jupiters. Our results also explain
why hot Jupiters had to stop migrating before falling on to their
parent stars. They also explain why, among pairs of resonant planets,
the outermost one is typically the most massive object. 

\vskip 12pt
\noindent{\bf Acknowledgments}

We wish to thank W. Kley, F. Masset, R.P. Nelson and A. Quillen for
discussions and suggestions. We are grateful to the French National
Programme of Planetary Science (PNP) for financial support. We
also wish to thank the anonymous reviewer for interesting suggestions
on the hot and warm Jupiters and the dust dynamics.

\bibliographystyle{./mnras}
\bibliography{./these2}

\begin{thebibliography}{35}
\expandafter\ifx\csname natexlab\endcsname\relax\def\natexlab#1{#1}\fi

\bibitem[{{Bacciotti} {et~al.}(2003){Bacciotti}, {Ray}, {Eisl{\"o}ffel},
  {Woitas}, {Solf}, {Mundt}, \& {Davis}}]{Bacciotti-etal-2003}
{Bacciotti} F., {Ray} T.~P., {Eisl{\"o}ffel} J., {Woitas} J., {Solf} J.,
  {Mundt} R.,  {Davis} C.~J. 2003, \apss, 287, 3

\bibitem[{{Beckwith}(1999)}]{Beckwith1999}
{Beckwith} S.~V.~W. 1999, in NATO ASIC Proc. 540: The Origin of Stars and
  Planetary Systems, ed. C.~J. {Lada}, N.~D. {Kylafis}, 579--+

\bibitem[{{Bergin} {et~al.}(2004){Bergin}, {Calvet}, {Sitko}, {Abgrall},
  {D'Alessio}, {Herczeg}, {Roueff}, {Qi}, {Lynch}, {Russell}, {Brafford}, \&
  {Perry}}]{Bergin-etal-2004}
{Bergin} E., {et~al.} 2004, \apjl, 614, L133

\bibitem[{{Bodenheimer} {et~al.}(2000){Bodenheimer}, {Hubickyj}, \&
  {Lissauer}}]{Bodenheimer-etal-2000}
{Bodenheimer} P., {Hubickyj} O.,  {Lissauer} J.~J. 2000, Icarus, 143, 2

\bibitem[{{Calvet} {et~al.}(2002){Calvet}, {D'Alessio}, {Hartmann}, {Wilner},
  {Walsh}, \& {Sitko}}]{Calvet-etal-2002}
{Calvet} N., {D'Alessio} P., {Hartmann} L., {Wilner} D., {Walsh} A.,  {Sitko}
  M. 2002, \apj, 568, 1008

\bibitem[{{Calvet} {et~al.}(2005){Calvet}, {D'Alessio}, {Watson},
  {Franco-Hern{\'a}ndez}, {Furlan}, {Green}, {Sutter}, {Forrest}, {Hartmann},
  {Uchida}, {Keller}, {Sargent}, {Najita}, {Herter}, {Barry}, \&
  {Hall}}]{Calvet-etal-2005}
{Calvet} N., {et~al.} 2005, \apjl, 630, L185

\bibitem[{{Coffey} {et~al.}(2004){Coffey}, {Bacciotti}, {Woitas}, {Ray}, \&
  {Eisl{\"o}ffel}}]{Coffey-etal-2004}
{Coffey} D., {Bacciotti} F., {Woitas} J., {Ray} T.~P.,  {Eisl{\"o}ffel} J.
  2004, \apj, 604, 758

\bibitem[{{Crida} {et~al.}(2006){Crida}, {Morbidelli}, \&
  {Masset}}]{Crida-etal-2006}
{Crida} A., {Morbidelli} A.,  {Masset} F. 2006, Icarus, 181, 587

\bibitem[{{Crida} {et~al.}(2007){Crida}, {Morbidelli}, \&
  {Masset}}]{Crida-etal-2007}
{Crida} A., {Morbidelli} A.,  {Masset} F. 2007, \aap, 461, 1173

\bibitem[{{Dullemond} {et~al.}(2001){Dullemond}, {Dominik}, \&
  {Natta}}]{Dullemond-etal-2001}
{Dullemond} C.~P., {Dominik} C.,  {Natta} A. 2001, \apj, 560, 957

\bibitem[{{Forrest} {et~al.}(2004){Forrest}, {Sargent}, {Furlan}, {D'Alessio},
  {Calvet}, {Hartmann}, {Uchida}, {Green}, {Watson}, {Chen}, {Kemper},
  {Keller}, {Sloan}, {Herter}, {Brandl}, {Houck}, {Barry}, {Hall}, {Morris},
  {Najita}, \& {Myers}}]{Forrest-etal-2004}
{Forrest} W.~J., {et~al.} 2004, \apjs, 154, 443

\bibitem[{{Goldreich} \& {Tremaine}(1979)}]{GT79}
{Goldreich} P. {Tremaine} S. 1979, \apj, 233, 857

\bibitem[{{Guillot} \& {Hueso}(2006)}]{GuillotHueso2006}
{Guillot} T. {Hueso} R. 2006, \mnras, 367, L47

\bibitem[{{Hayashi}(1981)}]{Hayashi1981}
{Hayashi} C. 1981, Progress of Theoretical Physics Supplement, 70, 35

\bibitem[{{Kley}(1999)}]{Kley1999}
{Kley} W. 1999, \mnras, 303, 696

\bibitem[{{Kuchner} \& {Lecar}(2002)}]{KuchnerLecar2002}
{Kuchner} M.~J. {Lecar} M. 2002, \apjl, 574, L87

\bibitem[{{Lin} \& {Papaloizou}(1979)}]{LinPapaloizou1979}
{Lin} D.~N.~C. {Papaloizou} J. 1979, \mnras, 186, 799

\bibitem[{{Lubow} \& {D'Angelo}(2006)}]{LubowDAngelo2006}
{Lubow} S.~H. {D'Angelo} G. 2006, \apj, 641, 526

\bibitem[{{Lynden-Bell} \& {Pringle}(1974)}]{LBP74}
{Lynden-Bell} D. {Pringle} J.~E. 1974, \mnras, 168, 603

\bibitem[{{Masset}(2000{\natexlab{a}})}]{FARGO}
{Masset} F. 2000{\natexlab{a}}, \aaps, 141, 165

\bibitem[{{Masset}(2000{\natexlab{b}})}]{FARGO2}
{Masset} F. 2000{\natexlab{b}}, in ASP Conf. Ser. 219: Disks, Planetesimals,
  and Planets, ed. G.~{Garz{\'o}n}, C.~{Eiroa}, D.~{de Winter},, T.~J.
  {Mahoney}, 75--80

\bibitem[{{Masset}(2001)}]{Masset2001}
{Masset} F.~S. 2001, \apj, 558, 453

\bibitem[{{Mayor} \& {Queloz}(1995)}]{MayorQueloz1995}
{Mayor} M. {Queloz} D. 1995, \nat, 378, 355

\bibitem[{{Paardekooper} \& {Mellema}(2006)}]{PaardekooperMellema2006_dust}
{Paardekooper} S.-J. {Mellema} G. 2006, \aap, 453, 1129

\bibitem[{{Pi{\'e}tu} {et~al.}(2006){Pi{\'e}tu}, {Dutrey}, {Guilloteau},
  {Chapillon}, \& {Pety}}]{Pietu-etal-2006}
{Pi{\'e}tu} V., {Dutrey} A., {Guilloteau} S., {Chapillon} E.,  {Pety} J. 2006,
  \aap, 460, L43

\bibitem[{{Pollack} {et~al.}(1996){Pollack}, {Hubickyj}, {Bodenheimer},
  {Lissauer}, {Podolak}, \& {Greenzweig}}]{Pollack-etal-1996}
{Pollack} J.~B., {Hubickyj} O., {Bodenheimer} P., {Lissauer} J.~J., {Podolak}
  M.,  {Greenzweig} Y. 1996, Icarus, 124, 62

\bibitem[{{Quillen} {et~al.}(2004){Quillen}, {Blackman}, {Frank}, \&
  {Varni{\`e}re}}]{Quillen-etal-2004}
{Quillen} A.~C., {Blackman} E.~G., {Frank} A.,  {Varni{\`e}re} P. 2004, \apjl,
  612, L137

\bibitem[{{Rice} {et~al.}(2006){Rice}, {Armitage}, {Wood}, \&
  {Lodato}}]{Rice-etal-2006}
{Rice} W.~K.~M., {Armitage} P.~J., {Wood} K.,  {Lodato} G. 2006, \mnras, in
  press

\bibitem[{{Rice} {et~al.}(2003){Rice}, {Wood}, {Armitage}, {Whitney}, \&
  {Bjorkman}}]{Rice-etal-2003}
{Rice} W.~K.~M., {Wood} K., {Armitage} P.~J., {Whitney} B.~A.,  {Bjorkman}
  J.~E. 2003, \mnras, 342, 79

\bibitem[{{Shakura} \& {Sunyaev}(1973)}]{ShakuraSunyaev1973}
{Shakura} N.~I. {Sunyaev} R.~A. 1973, \aap, 24, 337

\bibitem[{{Shu} {et~al.}(1997){Shu}, {Shang}, {Glassgold}, \&
  {Lee}}]{Shu-et-al-1977}
{Shu} F.~H., {Shang} H., {Glassgold} A.~E.,  {Lee} T. 1997, Science, 277, 1475

\bibitem[{{Varni{\`e}re} {et~al.}(2006){Varni{\`e}re}, {Blackman}, {Frank}, \&
  {Quillen}}]{Varniere-etal-2006}
{Varni{\`e}re} P., {Blackman} E.~G., {Frank} A.,  {Quillen} A.~C. 2006, \apj,
  640, 1110

\bibitem[{{Ward}(1991)}]{Ward1991}
{Ward} W.~R. 1991, in Lunar and Planetary Institute Conference Abstracts,
  1463--1464

\bibitem[{{Ward}(1997)}]{Ward1997}
{Ward} W.~R. 1997, Icarus, 126, 261

\bibitem[{{Weidenschilling}(1977)}]{Weidenschilling77}
{Weidenschilling} S.~J. 1977, \mnras, 180, 57

\end{thebibliography}

\end{document}